\documentclass[journal]{IEEEtran}
\usepackage{amsmath}
\usepackage[T1]{fontenc}
\usepackage{color}
\usepackage{babel}
\usepackage{array}
\usepackage{float}
\usepackage{multirow}
\usepackage{amsmath}
\usepackage{graphicx}
\usepackage[none]{hyphenat}
\usepackage[noadjust]{cite} 
\usepackage{booktabs} 
\usepackage{multirow}  

\ifCLASSINFOpdf
\else
\fi
\begin{document}

\title{Phasor Modelling Approaches and Simulation Guidelines of Voltage-Source Converters in Grid-Integration Studies}



\author{Vin\'{\i}cius~A.~Lacerda,
        Eduardo~Prieto~Araujo,~\IEEEmembership{Senior Member,~IEEE,}
        Marc~Cheah-Ma\~{n}e,~\IEEEmembership{Member,~IEEE,}
        and~Oriol~Gomis-Bellmunt,~\IEEEmembership{Fellow,~IEEE}
\thanks{V. A. Lacerda, E. Prieto-Araujo, M. Cheah-Ma\~{n}e and O. Gomis-Bellmunt  was with the Centre d'Innovacio Tecnol\`{o}gica en Convertidors Estatics i Accionaments Universitat Polit\`{e}cnica de Catalunya (CITCEA-UPC), Av Diagonal 647, H building, 2nd floor, Barcelona, Spain. e-mail: (vinicius.lacerda, eduardo.prieto-araujo, marc.cheah, oriol.gomis)@upc.edu.}
\thanks{This project has received funding from the European Union’s Horizon 2020 research and innovation programme under grant agreement No 883985 (POSYTYF project). The work of Oriol Gomis-Bellmunt is supported by the ICREA Academia program. Eduardo Prieto-Araujo and Marc~Cheah-Ma\~{n}e are Serra H\'{u}nter lecturers.}
}

\markboth{Journal of \LaTeX\ Class Files,~Vol.~1, No.~1, August~2021}%
{Shell \MakeLowercase{\textit{et al.}}: Bare Demo of IEEEtran.cls for IEEE Journals}

\maketitle

\begin{abstract}
This paper reviews Voltage-Source Converters (VSCs) EMT and Phasor models currently used to simulate converter-interfaced generation (CIG) and renewable energy resources integration to power systems. Several modelling guidelines and suitability analyses were provided based on a comprehensive comparative study among the models. Various studies were performed in a small system and a large system, modelled in Simulink. We address a gap related to the suitability of CIGs phasor models in studies where the boundary between electromagnetic and electromechanical transients overlap. An insightful analysis of the adequate simulation time step for each model and study is also provided, along with several simulation guidelines.

	

\end{abstract}

\begin{IEEEkeywords}
converter-interfaced generation (CIG), grid-integration studies, phasor model, phasor simulation, voltage-source converter.
\end{IEEEkeywords}

\IEEEpeerreviewmaketitle

\section{Introduction}
\IEEEPARstart{I}{n} power system studies, it is common to make simplifying assumptions to simulate systems in a limited computation time. In the past, the boundary between electromagnetic and electromechanical studies was clear, facilitating the use of these assumptions \cite{Kundur_2004}.

However, with the recent large integration of converter-interfaced generation (CIG) and widespread of power electronic interfaced technologies, power system dynamic behaviour is becoming progressively faster, making both modelling approaches to overlap \cite{Hatziargyriou_2021}. In this context, it is often unclear when it is appropriate to use EMT, Phasor Mode (PM) or other alternative simulation methods.

The significant research activity into the control and stability of power systems with large penetration of CIGs and HVDC systems is fostering the search for new simulation methods. The challenge is to have a sufficiently accurate model to capture the fast dynamics of power electronics components, but not computation-intensive to enable simulating large systems. One line of research focus on new modelling methods such as Dynamic Phasors \cite{Sanders_1991,Demiray_2006,Daryabak_2014}, Harmonic State-Space modelling \cite{Love_2008,Arrillaga_2008,Hwang_2012,Hwang_2015,Kwon_2017} Harmonic Phasor Domain \cite{Shu_2021} and frequency-dependent equivalents \cite{Milano_2019,D_Arco_2021}, that aim to model more precisely the fast dynamics of power electronics but keeping the computational cost lower than in EMT. Another line of research focuses on co-simulation methods \cite{Jalili_Marandi_2009,Zhang_2013,van_der_Meer_2015,Zadkhast_2021}, where the power grid is often modelled in PM or another low-frequency equivalent model while the converter or power electronic device is modelled in EMT.

However, little research has been made on the capabilities of full PM models to perform CIG integration studies when the boundary between electromagnetic and electromechanical domains overlap. While important recommendations and guidelines were recently available \cite{De_Carne_2019,Paolone_2020}, those are often high-level and based on the researchers' experience, and important questions still need to be addressed, such as the maximum simulation time step for each study, the use of PM models in short-circuit studies and the influence of harmonics in PM simulation.

Therefore, this paper reviews the models currently used to simulate CIG based on Voltage-Source Converters (VSC). It also analyses the suitability of PM models to several power system studies and provides simulation guidelines based on a comprehensive comparative study of EMT and PM models. The paper is organized as follows. Section~\ref{sec_conv_modelling} introduces the conventional VSC modelling in EMT. Section~\ref{sec_phasor_modelling} briefly describes phasor simulation and approximated VSC phasor models. Section~\ref{sec_methodology} presents the methodology to conduct the comparative analysis, including the systems simulated and the tests performed. The results are shown in Section~\ref{sec_results} along with punctual discussions. General discussion and simulation guidelines are provided in   
Section~\ref{sec_guidelines}. Finally, the conclusions are drawn in
Section~\ref{sec_conclusion}.


\section{VSC EMT simulation}\label{sec_conv_modelling}
Several simulation models of VSC were proposed, varying from detailed models, in the semiconductors domain, to high-level RMS models used in power flow and system planning studies \cite{Cigre_604,Khan_2017}. The proper choice will depend on the level of detail, the phenomena being analysed, and the time available for simulation. Three common types of VSC models are the switching model (SW), the average value model (AVG) and the phasor mode model (PM). The SW and AVG models run in EMT simulation. 

The EMT solution comprises a set of ordinary differential equations (ODEs) based on Kirchhoff’s Current and Voltage laws (KCL and KVL, respectively). This set of ODEs describes the circuit’s complete response when subjected to a voltage or current \cite{Watson_2018}. 
There are several numerical methods to solve ODEs. One largely used in power systems simulation is the one proposed by \cite{Dommel_1969}. The method was called EMTP (Electromagnetic Transients Program) and was based on the difference equations model obtained with the trapezoidal rule. 
Several well-known commercial and non-commercial programs are based, or use to some extent, the original EMTP and its improvements, such as ATP, PSCAD, EMTP-RV, RTDS and PowerFactory. After decades of validations and improvement, the EMTP method is the most generally accepted method to perform EMT simulations \cite{Watson_2018}. 

\subsection{EMT AVG model}
Several models have been proposed for the conventional two- and three-level VSCs. One widely-used model is the AVG model.
The AVG model neglects the converter's high-frequency switching and considers that the VSC output voltage is a linearly amplified form of the modulation index, which is nearly sinusoidal \cite{Weixing_Lu_2003}. Thus, the VSC electrical model on the AC side is simply a controlled voltage source. As a result, the AVG model focuses only on the power frequency component, which effectively transmits power to/from the grid.

\subsection{Control system}
The modulation index is defined by a control system. In this study, a classic hierarchical structure is presented. The inner loop regulated the positive and negative sequence output current in the $dq$ frame using the double synchronous reference frame (DSRF) \cite{Hong_Seok_Song_1999}. The outer loop regulated the AC power injected into the grid. Proportional-integral controllers (PI) of the form $G(s) \!=\! k_{p}(1\!+\!1/(\tau_{i}s))$ were used in both loops. 

\begin{figure}[ht]
	\vspace{-3pt}
	\centering
	\includegraphics[width=0.38\textwidth]{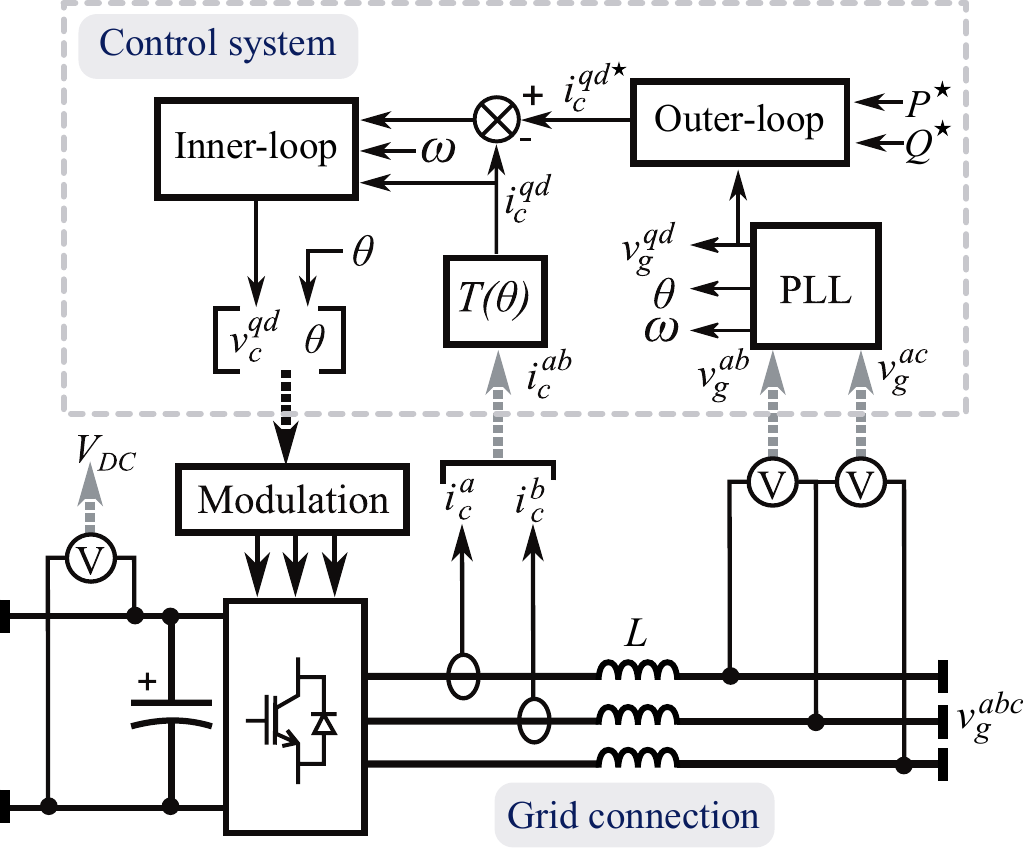}
	\vspace{-3pt}
	\caption{VSC diagram.}
		\vspace{-6pt}
	\label{fig_VSC_diagram}
\end{figure}

The converter output current in the $abc$ frame ($i_c^{abc}$) is calculated as

\begin{equation}
    v_c^{abc} - v_g^{abc} = R i_c^{abc} + L\dfrac{\mathrm{d}}{\mathrm{d}t}i_c^{abc}
    \label{eq_VSC_conv_abc}
\end{equation}

where $v_c^{abc}$ is the converter output voltage, $v_g^{abc}$ is the grid voltage and $R$ and $L$ are the equivalent resistance and inductance from the converter to the point of common coupling, respectively, as shown in Fig.~\ref{fig_VSC_diagram}.

Transforming \eqref{eq_VSC_conv_abc} to the $dq$ frame yields

\begin{equation}
    v_c^{qd} - v_g^{qd} = \begin{bmatrix}
R & L \dot{\theta}\\ 
-L \dot{\theta} & R
\end{bmatrix}
i_c^{qd} + L\dfrac{\mathrm{d}}{\mathrm{d}t}i_c^{qd}
    \label{eq_VSC_conv_dq}
\end{equation}

where $\dot{\theta}$ is the voltage phase time derivative (grid instantaneous angular frequency) and it is used in the Park transform to produce constant voltages and currents in the $dq$ frame.

The synchronized $\theta$ is obtained from the Phase-locked Loop (PLL), aligned with $v_q$ such that $v_d = 0$. In this study, the conventional PLL and tunning approach proposed by \cite{Se_Kyo_Chung_2000} was used. The PLL gains were
\begin{align}
	\zeta_{pll} &= 1/\sqrt{2} \\	
	k_{p,pll} &= 2 \zeta_{pll} \omega_{pll}/V_{g,pk} \\	
	\tau_{i,pll} &= 2 \zeta_{pll}/\omega_{pll}
\end{align}

where $\zeta_{pll}$ is the damping factor, $k_{p,pll}$ is the proportional gain, $\tau_{i,pll}$ is the integral time constant and $V_{g,pk}$ is the grid phase-to-ground peak voltage.

The inner-loop PI controller 
gains are tuned using the internal model control approach \cite{Harnefors_1998}, yielding
\begin{align}
	\tau_{i,c} = L/R \\	
	k_{p,c} = L \omega_c
\end{align}

where $\omega_c$ is the desired closed-loop bandwidth, $k_{p,c}$ is the proportional gain and $\tau_{i,c}$ is the integral time constant.

The outer-loop PI controllers are identical for both active and reactive power control, with gains tuned according to the modulus optimum criteria \cite{bajracharya2008}, yielding
\begin{align}
	\tau_{i,pq} &= 1/\omega_c\\
	k_{p,pq} &= \tau_{i,pq} \omega_{pq} \, 2/(3 V_{g,pk})	
\end{align}

where $\omega_{pq}$ is the desired closed-loop bandwidth, $k_{p,pq}$ is the proportional gain and $\tau_{i,pq}$ is the integral time constant.

\subsubsection{Frequency droop}
The conventional frequency droop without deadband is implemented, where the additional active power injected by the VSC into the grid ($\Delta P$) is proportional ($k_{droop,f}$) to the frequency deviation ($\Delta f$):

\begin{equation}
\Delta P = k_{droop,f} \Delta f 	 
\end{equation}

where $\Delta f$ is obtained by low-pass filtering $\dot{\theta}$, which is estimated by the PLL.

\subsubsection{Voltage droop}
Similarly, the conventional voltage droop is implemented, where the additional reactive power injected by the VSC into the grid ($\Delta Q$) is proportional ($k_{droop,v}$) to the voltage deviation ($\Delta |v|$):

\begin{equation}
	\Delta Q = k_{droop,v} \Delta |v| 	 
\end{equation}

where $\Delta |v|$ is obtained by low-pass filtering $v_g^q$.

\subsubsection{Low voltage ride through (LVRT)}
When the grid voltage drops below a certain level ($v_{g,min}$), the VSC is often required to inject reactive current to support to grid. This characteristic is implemented following a LVRT curve, where the reactive current reference is defined by
 
\begin{equation}
	i_c^{d\star} = k_{lvrt} (v_{g,min} - v_g^q)
\end{equation}

When the LVRT characteristic is activated, during the low-voltage condition, the current reference $i_c^{d\star}$ is defined only by the LVRT characteristic and not by the reactive power control loop, to avoid incurring in two conflicting control objectives. $i_c^{d\star}$ is saturated if exceeds the limit $i_{lvrt,max}$, which defined as a parameter of the LVRT characteristic.

\subsubsection{Negative sequence control}
A sequence extraction technique is required to implement the DSRF, which can be performed by notch filtering, high-pass filtering, or signal cancellation techniques such as the Delayed Signal Cancellation (DSC). In this study, the DSC was used due to its simplicity, despite its non-ideal performance during off-nominal grid conditions. As both positive and negative sequence components were available, the control approach adopted was to have the PLL aligned with the positive sequence voltage ($v_g^{q+}$) and to define negative sequence current references $i_c^{q-\star} = 0$ and $i_c^{d-\star} = 0$.



\section{Phasor simulation}\label{sec_phasor_modelling}
In typical transient stability studies, it is assumed that the power system frequency remains close to the nominal value (50~Hz or 60~Hz). Thus, phasor simulation aims to capture only the slow dynamics of the power grids, such as the electromechanical transients, with time constants generally bigger than 100 ms. In this mode, the grid differential equations resulting from the interaction of R, L, and C elements are substituted by algebraic equations. Also, the distributed models of transmission lines and cables are substituted by lumped models. Thus, the system voltages and currents are given by
\begin{equation}
    \mathbf{V} = \mathbf{Z} \mathbf{I}
\end{equation}

where $\mathbf{V}$ and $\mathbf{I}$ are the complex-valued vectors of node voltages and current injections, and $\mathbf{Z}$ is the impedance matrix.

As the phasors are assumed to be rotating at nominal angular speed, voltages and currents have their dynamics around 0 Hz, instead of 50 or 60 Hz. This allows to dramatically increase the simulation time step and consequentially the simulation speed. Depending on the time step used and on the phenomena being simulated, the simulation speed can be increased in one to three orders of magnitude compared to detailed EMT simulation, which is especially interesting when simulating large networks for long periods. However, the speed-up provided by phasor simulation must be used with care because several EMT phenomena are entirely neglected in phasor simulation, which can result in overestimated or underestimated values.

\begin{figure*}[t]
	\vspace{-3pt}
	\centering
	\includegraphics[width=0.93\textwidth]{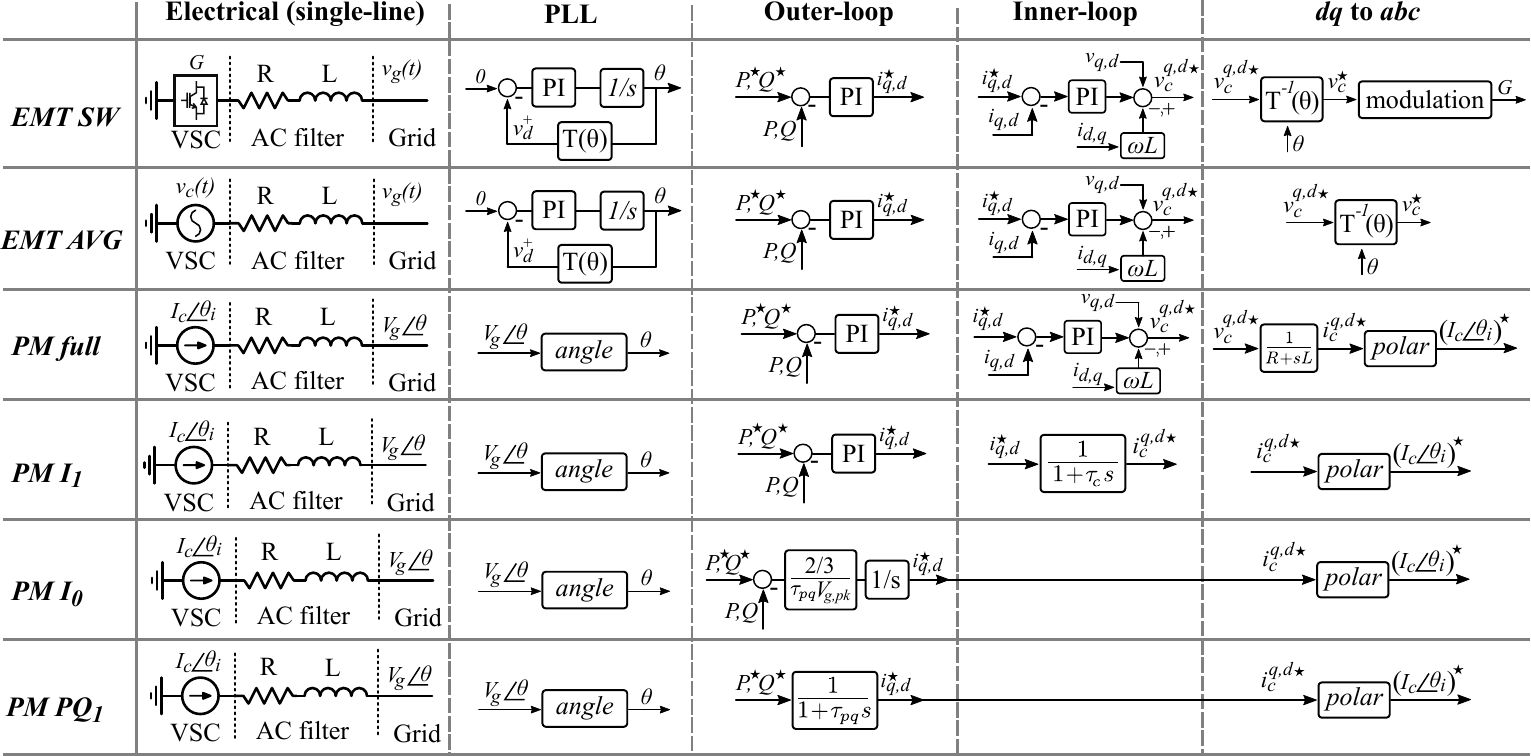}
	\vspace{-3pt}
	\caption{Summary of VSC EMT and PM models.}
	\label{fig_summary_VSCPMmodels}
\end{figure*}

In phasor simulation, the VSCs are represented as current sources with magnitude and angle defined by the control system, as depicted in Fig.\ref{fig_summary_VSCPMmodels}. As phasor simulation aims mainly to speed up and simplify simulations, several approximations can be performed in the VSC control system to allow bigger time steps to be used. In this study, four approaches were compared, which are described next and summarized in Fig.~\ref{fig_summary_VSCPMmodels}.

The conventional VSC model is composed of five parts: the electrical part, the transformation from \textit{abc} to \textit{dq}, the outer control loop, the inner control loop and the transformation from \textit{dq} to \textit{abc}. The electrical part and the transformation from \textit{abc} to \textit{dq} frame are the same amongst all phasor models. As in phasor simulation the voltage phase is readily available, no PLL is needed. To have $v_d^+ = 0$ similar to the EMT case, the angle ($\theta$) of the grid positive sequence voltage is taken and the voltages and currents in the $abc$ frame are rotated to this angle. Positive and negative sequences are calculated using Fortescue's theorem. Conversely, opposite rotation is performed when transforming from the \textit{dq} frame back to the \textit{abc} frame before passing the reference to the current source.

\subsection{Phasor model full}
In this model, the VSC inner and outer control loops are the same as those used in EMT. Thus, it completely represents the dynamics of the control system. As the output of the inner control is the reference voltage and the electrical part is a current source, the reference voltage is transformed in reference current using the electrical dynamics up to the point where the voltage is being measured. This approach has the benefit of modelling also the dynamics of the AC side connection, which is normally neglected in phasor simulation. In this study, an RL filter was used, so the dynamics are the ones described by the Laplace transform of \eqref{eq_VSC_conv_dq}
\begin{align}
	I_c^{d+\star} &=  (v_g^{d+} - v_c^{d+\star} + \omega_g L I_c^{q+})/(R + sL) \\
	I_c^{q+\star} &=  (v_g^{q+} - v_c^{q+\star} - \omega_g L I_c^{d+})/(R + sL) \\
	I_c^{d-\star} &=  (v_g^{d-} - v_c^{d-\star} - \omega_g L I_c^{q-})/(R + sL) \\
	I_c^{q-\star} &=  (v_g^{q-} - v_c^{q-\star} + \omega_g L I_c^{d-})/(R + sL) \\
	\label{eq_VSC_conv_dq_freq}
\end{align}

Using this approach, the complete control structure used in EMT simulations is preserved, which facilitates comparison between PM and EMT.

\subsection{Phasor model $I_1$} 
In this model, the VSC inner control is substituted by a first order transfer function with time constant $\tau_c$, defined to match the original closed-loop bandwidth:  

\begin{equation}
	\tau_c = \dfrac{1}{\omega_c}
\end{equation}

With this approximation, the DSRF closed-control structure is removed, including the cross-coupling term $\omega_gL$. As a consequence, the inner-loop output becomes the low-pass filtered ($\tau_c$) current reference defined by the outer-loop control, which is rotated back to the \textit{abc} frame and sent to the current source. 

\subsection{Phasor model $I_0$}
In this model, the VSC inner control is completely removed. As the inner control is the fastest dynamics of the VSC (when neglecting the switching), removing this loop allows bigger time steps to be used, which greatly speeds up simulations. However, it has the drawback of neglecting completely the dynamics of the current-control. Thus, it should be used in studies where these dynamics are not especially relevant.

Moreover, neglecting the time constant of the inner loop also affects the outer-loop. To maintain the outer-loop time constant equal to the original closed-loop we define:

\begin{equation}
	C_{pq}(s) = \dfrac{G_{pq}(s)}{1 + G_{pq}(s)} = \dfrac{1}{1 + \tau_{pq}s}
	\label{eq_PM_I0_1}
\end{equation}
where $\tau_{pq}$ is the original closed-loop time constant, $C_{pq}(s)$ is the new close-loop transfer function and $G_{pq}(s)$ is the new open-loop transfer function, defined as:

\begin{equation}
	G_{pq}(s) = K_{pq}(s)\dfrac{3}{2}v_g^q
\end{equation}
where $K_{pq}(s)$ is the new controller transfer function. Solving \eqref{eq_PM_I0_1} for $K_{pq}(s)$ yields:

\begin{equation}
	K_{pq}(s) = \dfrac{2}{3v_g^q} \dfrac{1}{\tau_{pq}s}.
\end{equation}

Thus, to have the same time constant, the original PI controller is substituted by an integrator with a gain equal to $2/(3v_g^q \tau_{pq})$. As $v_g^q$ is variable, it is often useful to consider a static gain by using $V_{g,pk} \! \approx \! v_g^q$. Note that in this structure, the signal sent to the integrator is the error between the reference and the measured power output, hence a closed-loop.
     
\subsection{Phasor model $PQ_1$}
Further simplification can be achieved if the outer-loop is substituted by a first-order transfer function with the same time constant ($\tau_{pq}$) as the original loop, similar to PM $I_1$:

\begin{equation}
	\tau_{pq} = \dfrac{1}{\omega_{pq}}
\end{equation}
   
With this approximation, the outer loop becomes an open-loop where the output current reference is defined by 

\begin{align}
	i^{q+\star} = \dfrac{2}{3} \dfrac{P^\star}{v_g^{q+}}, \qquad 
	i^{d+\star} = \dfrac{2}{3} \dfrac{Q^\star}{v_g^{q+}}
\end{align}
where $P^\star$ and $Q^\star$ are the active and reactive power references, respectively.

Next, the tests performed to assess the VSC models are described, followed by a discussion about their suitability.


\section{Methodology}\label{sec_methodology}
\begin{figure}[ht]
	\vspace{-3pt}
	\centering
	\includegraphics[width=0.41\textwidth]{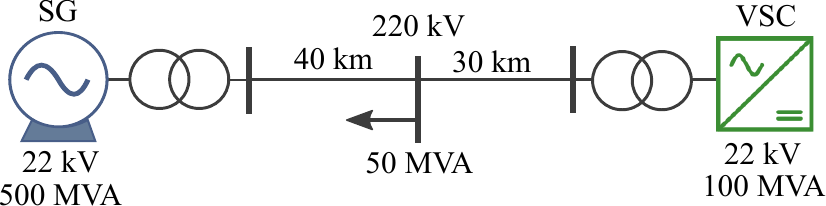}
	\vspace{-3pt}
	\caption{Small system single-line diagram.}
	\label{fig_small_system_diagram}
\end{figure}

In order to assess the aforementioned models, various studies were performed in a small system and a large system, modelled in Simulink. The small system was composed of a synchronous generator, a transmission line and a VSC. The large system was composed of four synchronous generators, eight transmission lines and two VSCs. The large system was inspired on the European HV transmission system in \cite{Cigre_575}, with a few adaptations. Both systems were simulated using a fixed time step and the Euler method (algorithm \textit{ode1} in Matlab). The single-line diagrams of the both systems are depicted in Fig.~\ref{fig_small_system_diagram} and Fig.~\ref{fig_large_system_diagram} and their parameters are summarized in Table~\ref{tab_SG_large_system}, in the Appendix.

\begin{figure*}[t]
	\vspace{-3pt}
	\centering
	\includegraphics[width=0.73\textwidth]{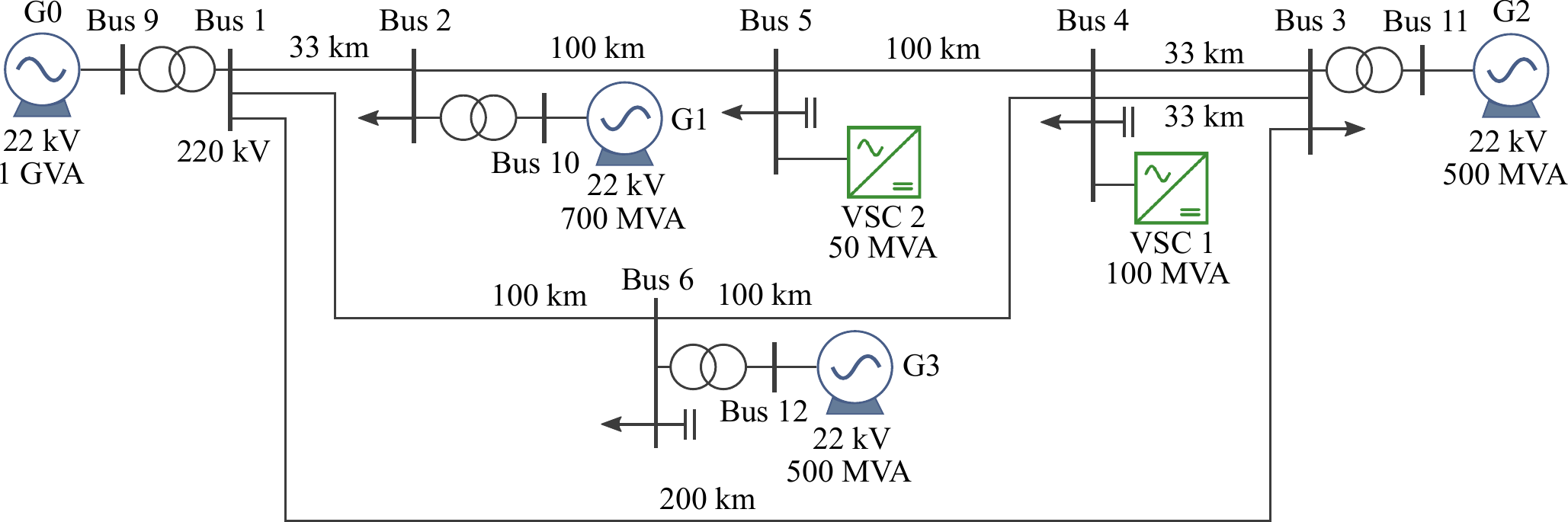}
	\vspace{-3pt}
	\caption{Large system single-line diagram. Modified from \cite{Cigre_575}.}
	\label{fig_large_system_diagram}
\end{figure*}

For all of the studies, the EMT AVG model running at 5~$\mu$s was defined as the reference model, while the phasor models and the AVG EMT model running at bigger time steps were compared against the reference model.

Three key aspects were analysed for each model: i) Precision, ii) Minimum simulation time step and iii) Total execution time.

The precision was quantified using the Root-Mean-Square Error (RMSE) between the variables simulated using each model and the variables simulated using the reference model:

\begin{equation}
RMSE = \sqrt{\dfrac{1}{N}\sum_{k=1}^{N}\big(x_{ref}(k)-\hat{x}(k)\big)^2}
\end{equation}

where $k$ is the sample, $N$ is the time window length, $x_{ref}$ is a variable in the reference model and $\hat{x}$ is the same variable in the model being assessed. In this study, the RMSE was divided by a proper base in each study, to present the results in a meaningful scale.

The minimum simulation time step was defined as the time step necessary to capture all the system dynamics and phenomena being analysed, which was assessed using the precision.

The total execution time was defined as the time necessary to run each study.  

Table~\ref{tab_tests_performed} summarizes the tests performed in both systems.

\begin{table}[]
\centering
\caption{Tests performed}
\label{tab_tests_performed}
\begin{tabular}{@{}lccc@{}}
\toprule
                            & Small system  \hspace{-6pt} & Large system \hspace{-12pt} & Variables    \\ \midrule
Setpoint tracking           & $\bullet$     &               & $P_{ac}$, $i_{dq}$           \\
Harmonics                   & $\bullet$     &               & $i_{dq}$, $v_{dq}$ \\
Frequency/Voltage deviation \hspace{-13pt} & $\bullet$     &      & $f$,$V_{abc}$      \\
Symmetrical faults          & $\bullet$     &      & $i_{dq}$, $v_{dq}$, $T_e$, $\omega_m$ \\ 
Asymmetrical faults         & $\bullet$     &  $\bullet$    &  $i_{dq}$, $v_{dq}$, $T_e$, $\omega_m$ \\ 
Loss of generation          &               &  $\bullet$    &  $i_{dq}$, $v_{dq}$, $T_e$, $\omega_m$ \\
Line outage                 &               &  $\bullet$    &   $i_{dq}$, $v_{dq}$, $T_e$, $\omega_m$ \\ \midrule 
Simulation time steps       & \multicolumn{3}{c}{from 5 $\mu s$ to 12000 $\mu s$} \\ \bottomrule
\end{tabular}
\end{table}

\section{Simulation results} \label{sec_results}
This section presents the simulation results, followed by brief punctual discussions. 
Due to the extensive number of tests, only a limited number of cases and variables are presented, which are representative of each test.

\subsection{Small system}

\subsubsection{Setpoint tracking}
\begin{figure}[htb]
	\centering
\includegraphics[width=0.46\textwidth, trim={0 15pt 0 0},clip]{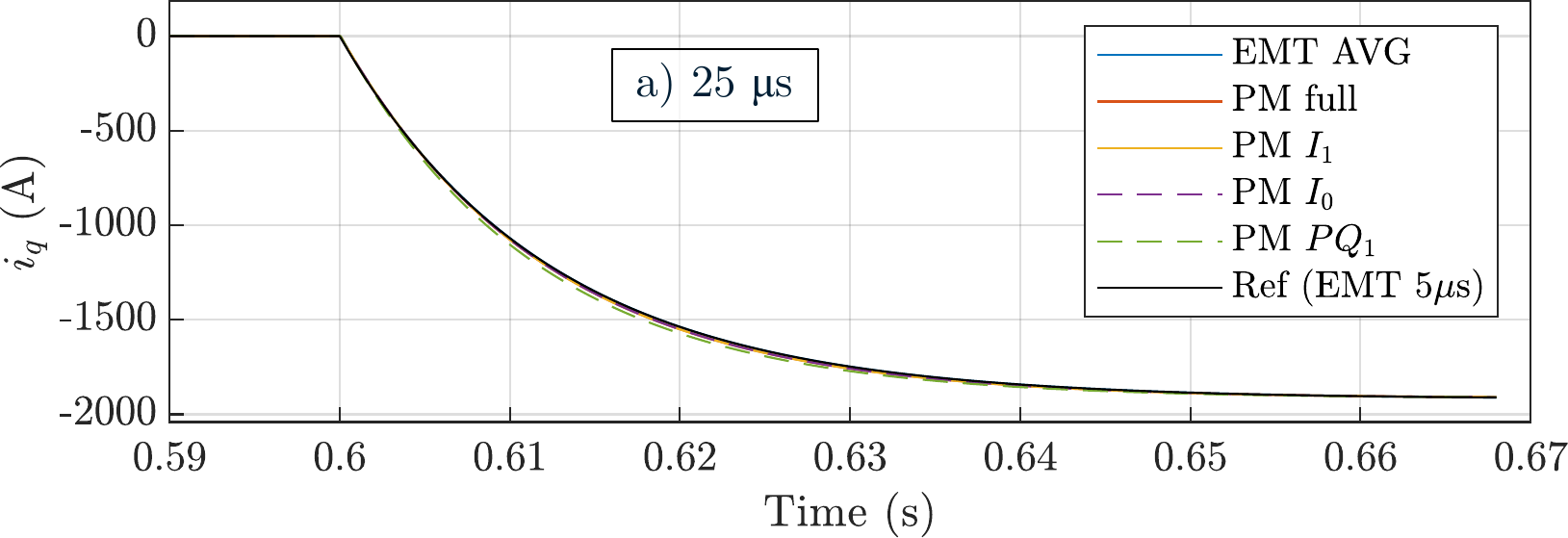}\vspace{2pt}
\includegraphics[width=0.46\textwidth, trim={0 15pt 0 0},clip]{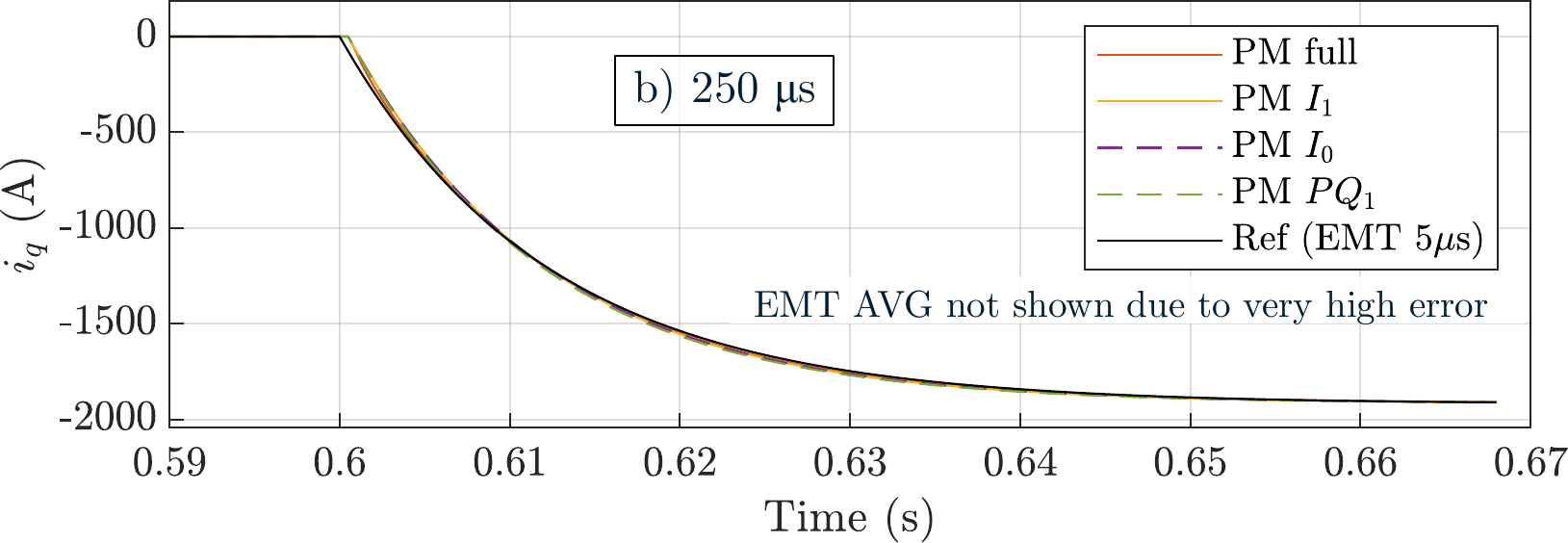}\vspace{2pt}	
\includegraphics[width=0.46\textwidth]{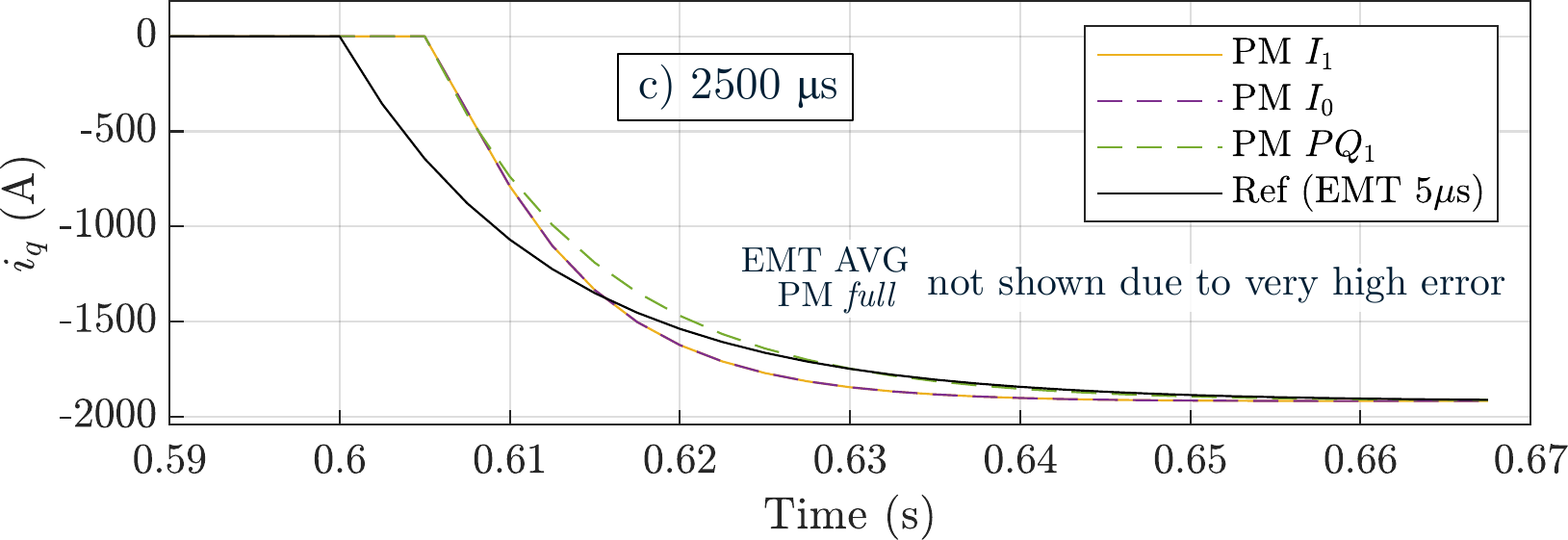}\vspace{-5pt}
\caption{Test 1 -- $i^{q+}$ setpoint tracking simulated by each model at a) $25 \,\mu$s, b) $250 \,\mu$s and c) $2500 \,\mu$s.}
\label{fig_test1_iqp_25us_250us_2500us}
\end{figure}

In this test, active and reactive power setpoints were set to 50~MW and 30~Mvar at $t \!=\! 0.6$~s and $t \!=\! 0.8$~s, respectively. 
Negative-sequence control and LVRT were disabled during this test, as well as frequency and voltage droop. Fig.~\ref{fig_test1_iqp_25us_250us_2500us} presents the converter currents for all models with a time step of $25 \,\mu$s, $250 \,\mu$s and $2500 \,\mu$s, respectively.

\begin{figure}[ht]
	\vspace{-3pt}
	\centering
	\includegraphics[width=0.46\textwidth]{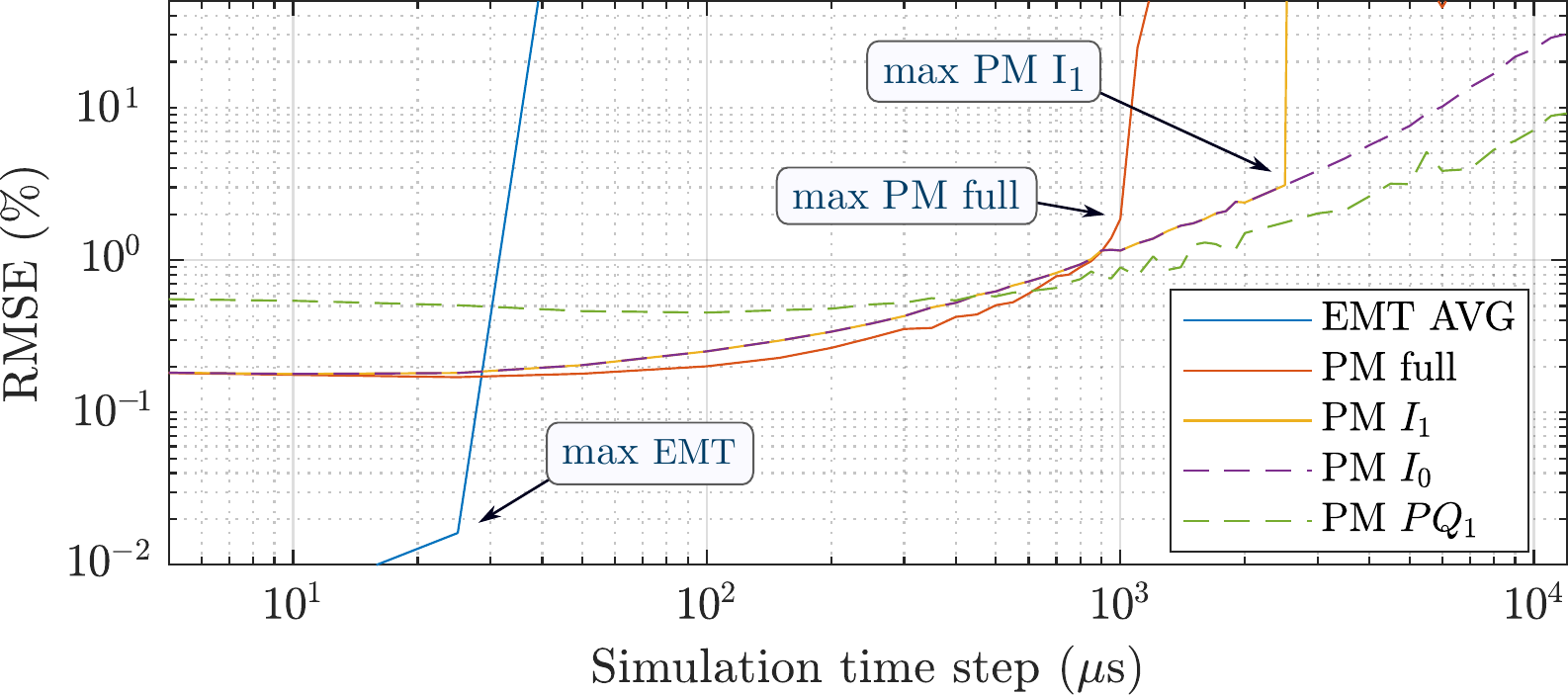}
	\vspace{-3pt}
	\caption{Test 1 -- $P_{ac}$ setpoint tracking simulated by EMT and PM models RMSE for several simulation time steps.}
	\label{fig_RMSE_iqp_test1}
\end{figure}

As can be observed, for smaller time steps, both EMT and PM models presented very similar results. However, as the simulation time step increases, the models that represent faster dynamics starts to deviate from the reference until they completely loss precision. This can be better visualized by varying the simulation time step between $5 \,\mu$s and $12000 \,\mu$s, as shown in Fig.~\ref{fig_RMSE_iqp_test1}. 

\begin{figure}[htb]
	\centering
	\includegraphics[width=0.46\textwidth, trim={0 17pt 0 0},clip]{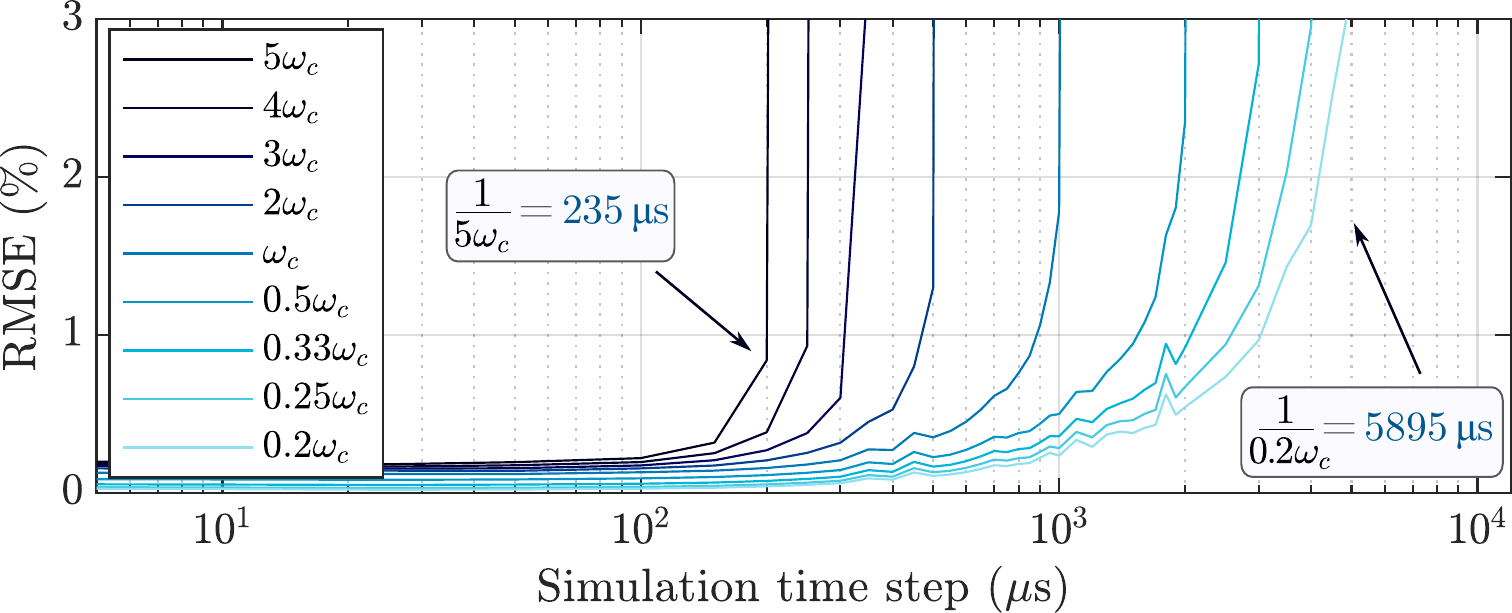}\vspace{1pt}
	\includegraphics[width=0.46\textwidth]{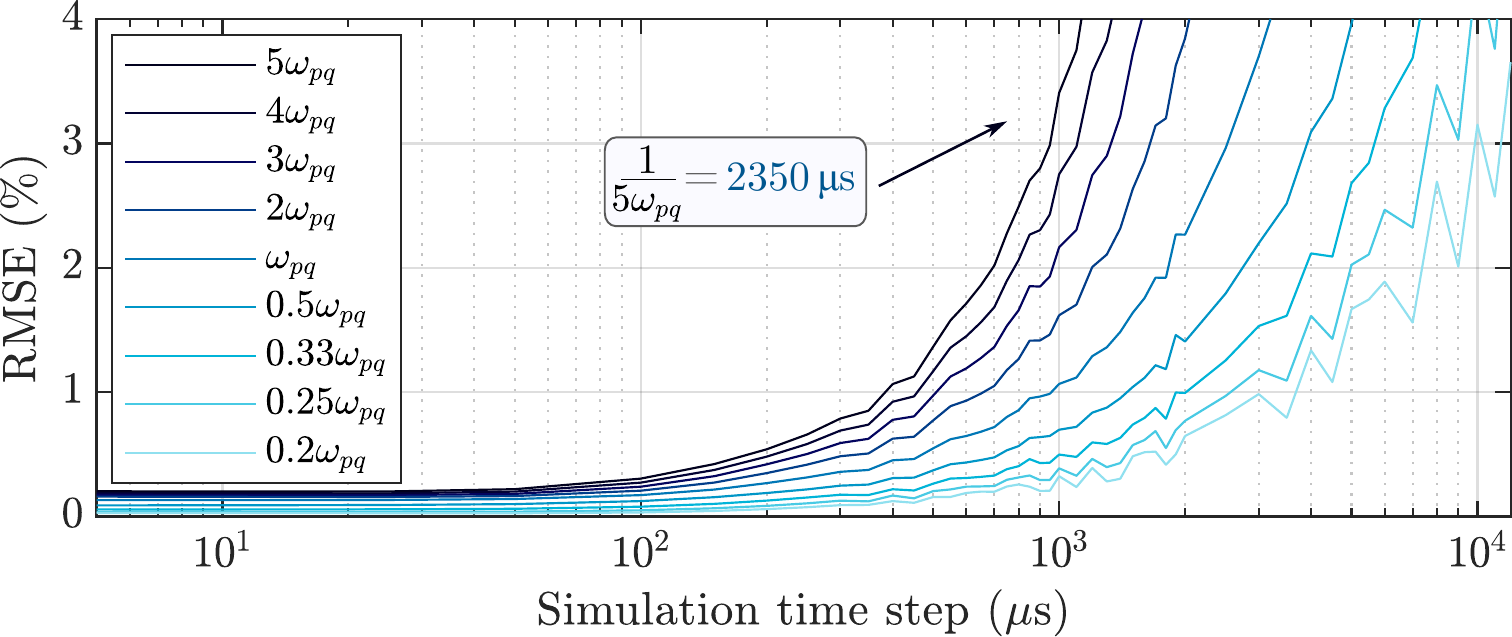}\vspace{-5pt}
	\caption{Test 1 -- $P_{ac}$ RMSE for several time steps and power control bandwidths. a) \textit{PM full} and b) \textit{PM} $I_0$ model.}
	\label{fig_wcs_test1_QEMT_Iref}
\end{figure}

When the simulation time step gets closer to the model's fastest dynamics, the error increases dramatically. In the system of Fig.~\ref{fig_small_system_diagram}, the fastest dynamics are related to the VSC current control loop, which have a time constant equal to $1/\omega_c = 1.18$~ms. This is more evident in Fig.~\ref{fig_wcs_test1_QEMT_Iref}a that shows the \textit{PM full} model error for several current control bandwidths. As the bandwidth increases, the maximum simulation time step to achieve low errors reduces proportionally. However, much higher time steps can be used (Fig.~\ref{fig_wcs_test1_QEMT_Iref}b) when performing the same test with the \textit{PM} $I_0$ model, as it neglects the current control loop and the fastest dynamics becomes the outer loop.  

As can be observed in this test, the PM $PQ_1$ model is worse than the other models for small time steps but its precision gets similar to the others for bigger time steps. Although the PM $I_1$ and $I_0$ approximate or neglect the current loop, they presented low error because their dynamics are smoothed by the outer loop, which is slower.

\subsubsection{Frequency/voltage deviation}
In this test, a 150 MW, 20 Mvar load is connected to the system at $t \!=\! 1\,s$, dropping the system frequency to around 49 Hz. The VSC frequency and voltage droops were enabled during this test.

\begin{figure}[ht]
	\vspace{-3pt}
	\centering
	\includegraphics[width=0.46\textwidth]{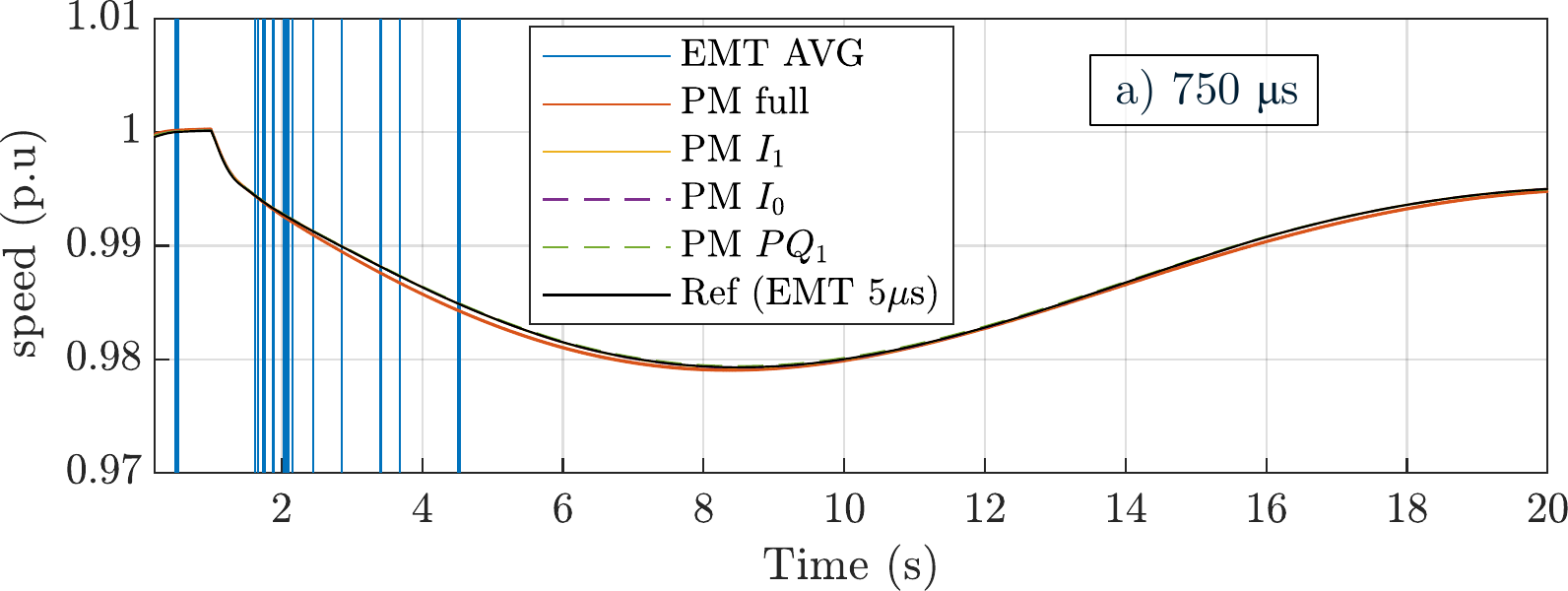}
	\vspace{-3pt}
	\caption{Test 2 -- $\omega_{m}$ simulated by each model at $750 \,\mu$s.}
	\label{fig_test2_wm_750us}
\end{figure}

\begin{figure}[ht]
	\vspace{-3pt}
	\centering
	\includegraphics[width=0.46\textwidth]{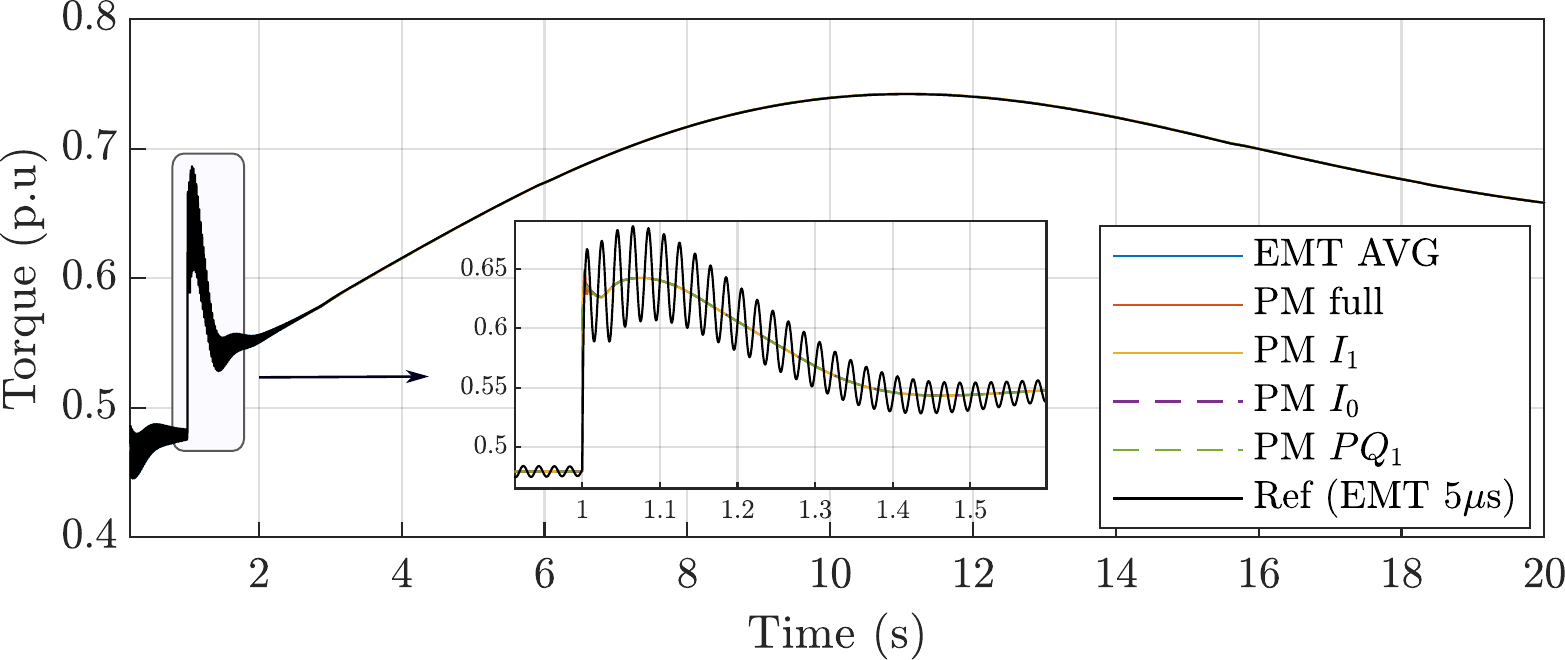}
	\vspace{-3pt}
	\caption{Test 2 -- $Te$ simulated by each model at $25 \,\mu$s.}
	\label{fig_test2_Te_25us_comzoom}
\end{figure}

As this test has slower dynamics, a few differences are observed between EMT and PM models, as can be observed in the SG speed in Fig.~\ref{fig_test2_wm_750us}. Due to the generators inertia, faster dynamics are mechanically ``low pass filtered'', which explains why the SG speed matches accurately even thought the electrical torque is oscillatory in EMT and constant in PM, as shown in Fig.~\ref{fig_test2_Te_25us_comzoom}. 

\begin{figure}[ht]
 	\vspace{-3pt}
 	\centering
 	\includegraphics[width=0.46\textwidth]{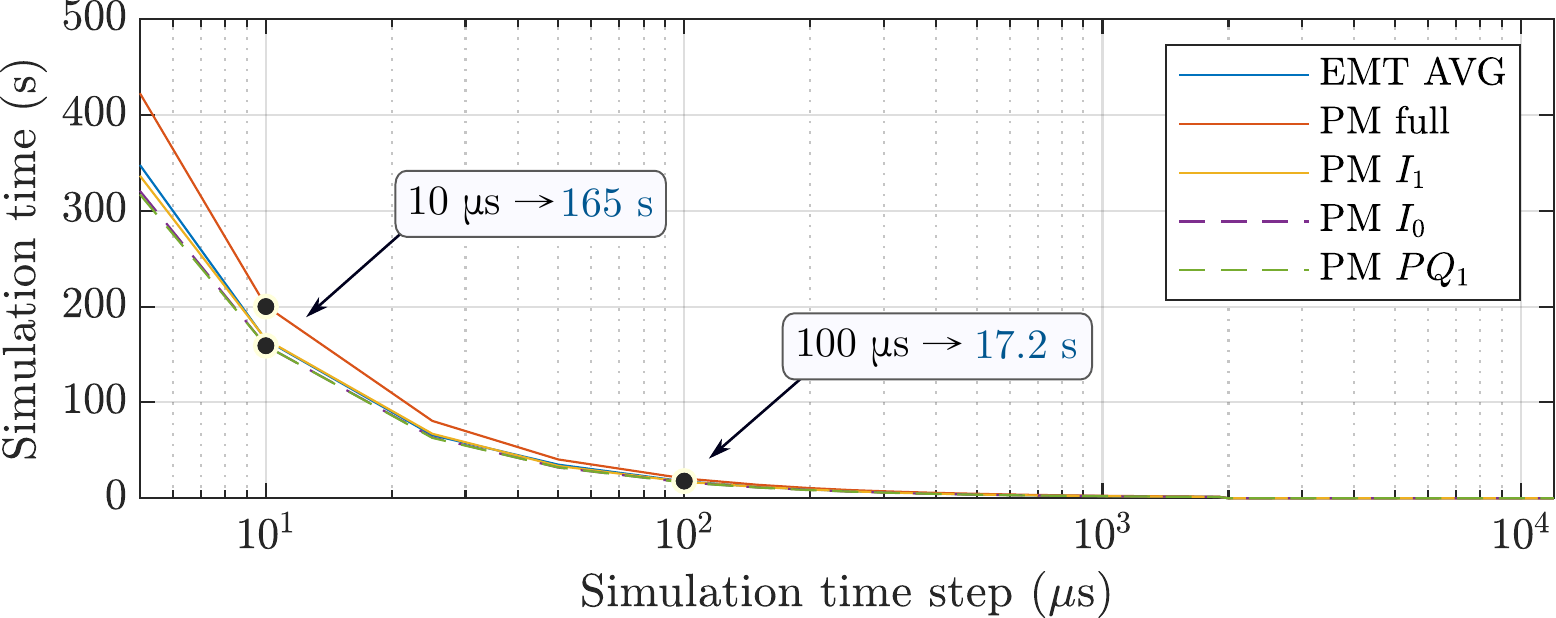}
 	\vspace{-3pt}
 	\caption{EMT and PM models simulation times for several time steps.}
 	\label{fig_simulation_time_test2}
 \end{figure}

The total execution time for each time step is shown in Fig.~\ref{fig_simulation_time_test2}. As can be observed in Fig.~\ref{fig_simulation_time_test2}, there is an inverse relationship between time step and execution time. For example, the average simulation time using a 10~$\mu$s step was 165~s while using a 100~$\mu$s step resulted in a average simulation time of 17.2~s, which is almost 10 times lower. If the step was further increased to 1000~$\mu$s, the simulation time was 2.2~s, 75 times faster than using 10~$\mu$s.

  \subsubsection{Symmetrical fault}
In this test, a 7~$\Omega$ three-phase fault was applied to the system at $t \!=\! 2.5\,s$, lasting for 300 ms. The LVRT was enabled during this test. 

\begin{figure}[ht]
	\vspace{-3pt}
	\centering
	\includegraphics[width=0.46\textwidth]{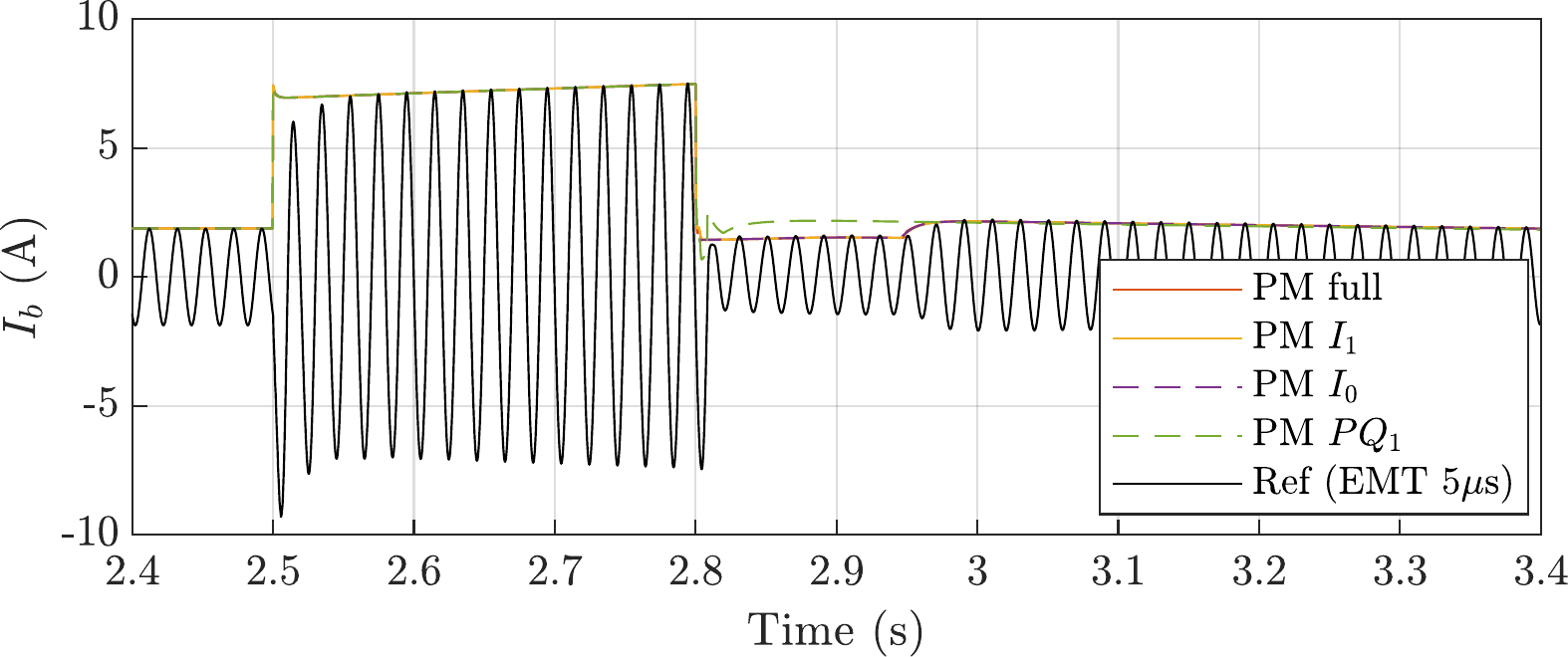}
	\vspace{-3pt}
	\caption{Test 3 -- $I_b$ simulated by each model at $350 \,\mu$s.}
	\label{fig_test3_Ib_350us}
\end{figure}

\begin{figure}[ht]
	\vspace{-3pt}
	\centering
	\includegraphics[width=0.46\textwidth]{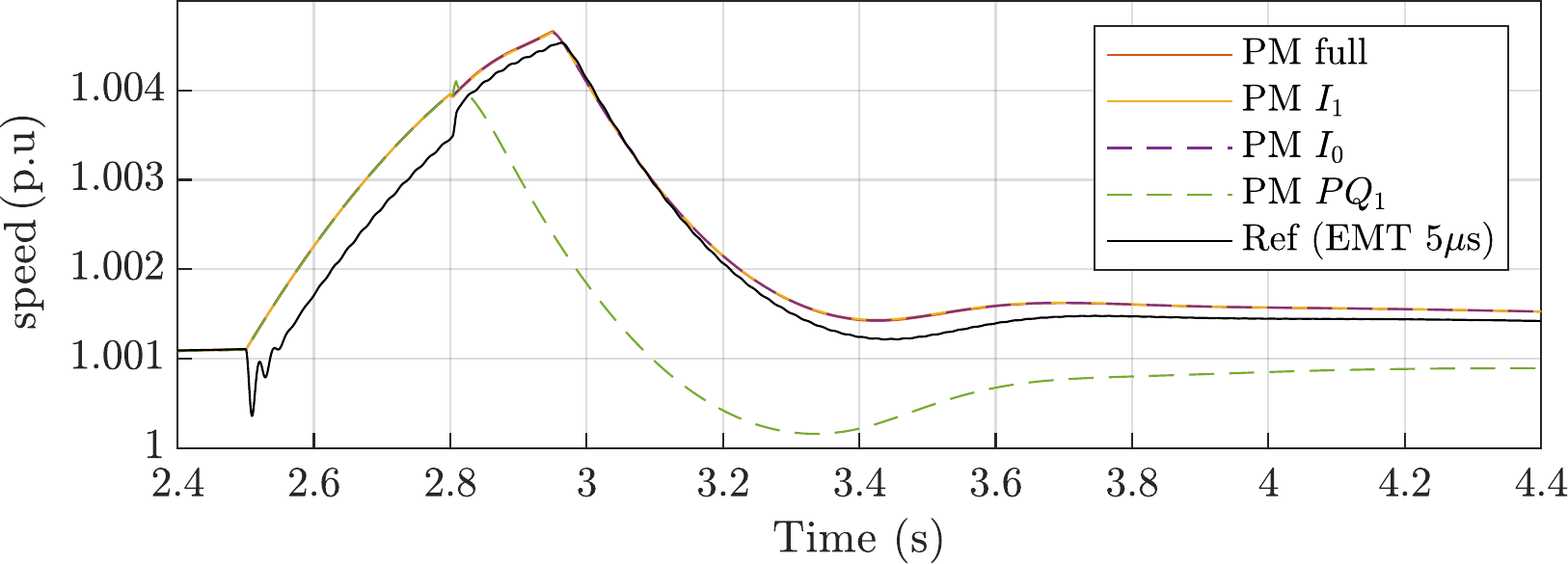}
	\vspace{-3pt}
	\caption{Test 3 -- $\omega_m$ simulated by each model at $600 \,\mu$s.}
	\label{fig_test3_wm_600us}
\end{figure}

Fig.~\ref{fig_test3_Ib_350us} shows the current at phase B. As can be observed in Fig.~\ref{fig_test3_Ib_350us}, PM models matched the EMT model closely in steady-state. After the transient, the PM models took only a few cycles to match the EMT model. The PM $PQ_1$ showed a slightly bigger deviation because as it has no PI controller in the outer loop, no saturation effects could be implemented, making the VSC injects more power just after the transient. This also affected the generator's speed, showed in Fig.~\ref{fig_test3_wm_600us}

 \begin{figure}[ht]
	\vspace{-3pt}
	\centering
	\includegraphics[width=0.46\textwidth]{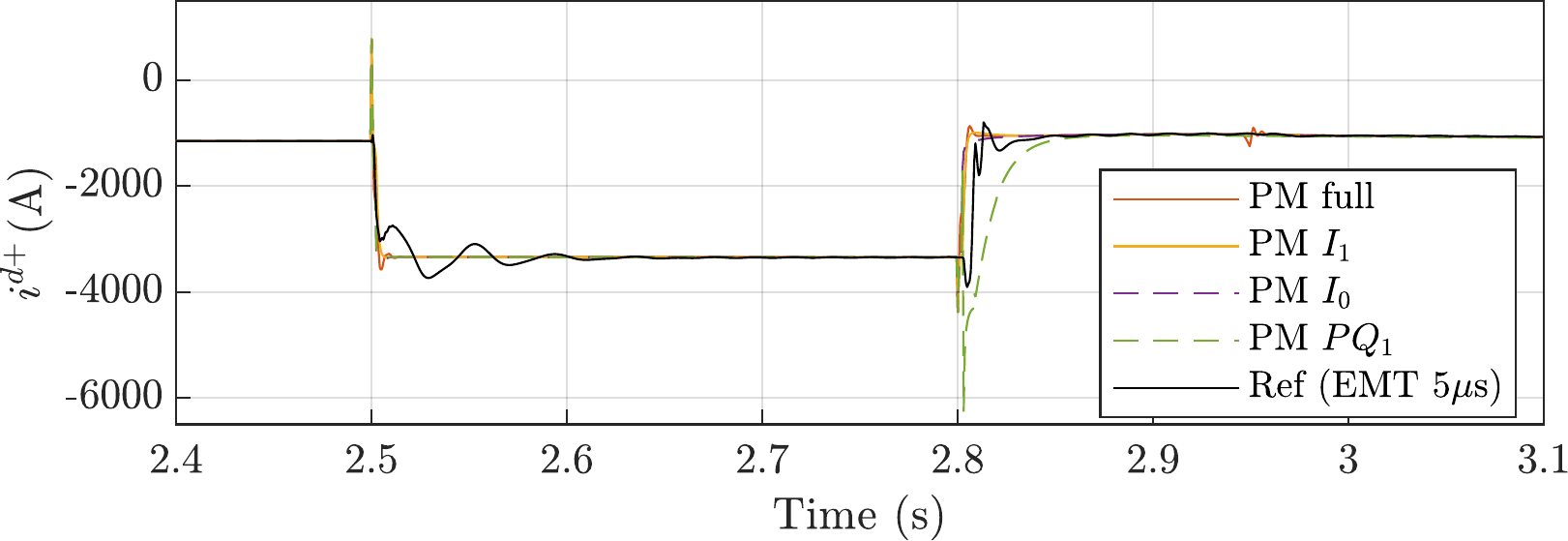}
	\vspace{-3pt}
	\caption{Test 3 -- $i^{d+}$ simulated by each model at $600 \,\mu$s.}
	\label{fig_test3_idp_600us}
\end{figure}

  \subsubsection{Asymmetrical fault}
In this test, a 10~$\Omega$ mono-phase fault at phase B was applied to the system at $t \!=\! 2.5\,s$, lasting for 300 ms. The negative-sequence control was enabled during this test.

\begin{figure}[ht]
	\vspace{-3pt}
	\centering
	\includegraphics[width=0.46\textwidth]{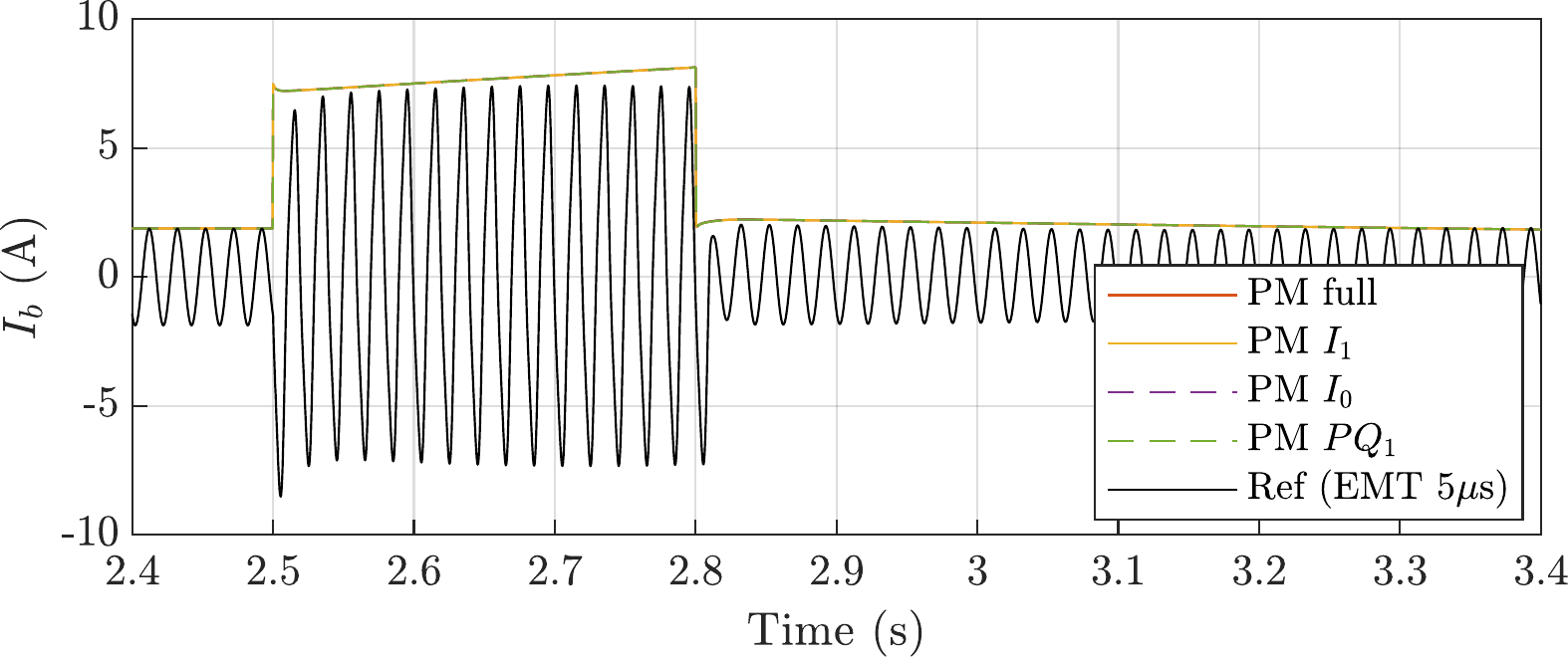}
	\vspace{-3pt}
	\caption{Test 4 -- $I_b$ simulated by each model at $350 \,\mu$s.}
	\label{fig_test4_Ib_350us}
\end{figure}

Fig.~\ref{fig_test4_Ib_350us} shows the current at phase B. Comparing Fig.~\ref{fig_test4_Ib_350us} with Fig.~\ref{fig_test3_Ib_350us} it can be observed that PM models presented higher error during the transient for asymmetric faults, reducing, however, after 100~ms.

\begin{figure}[ht]
	\vspace{-3pt}
	\centering
	\includegraphics[width=0.46\textwidth]{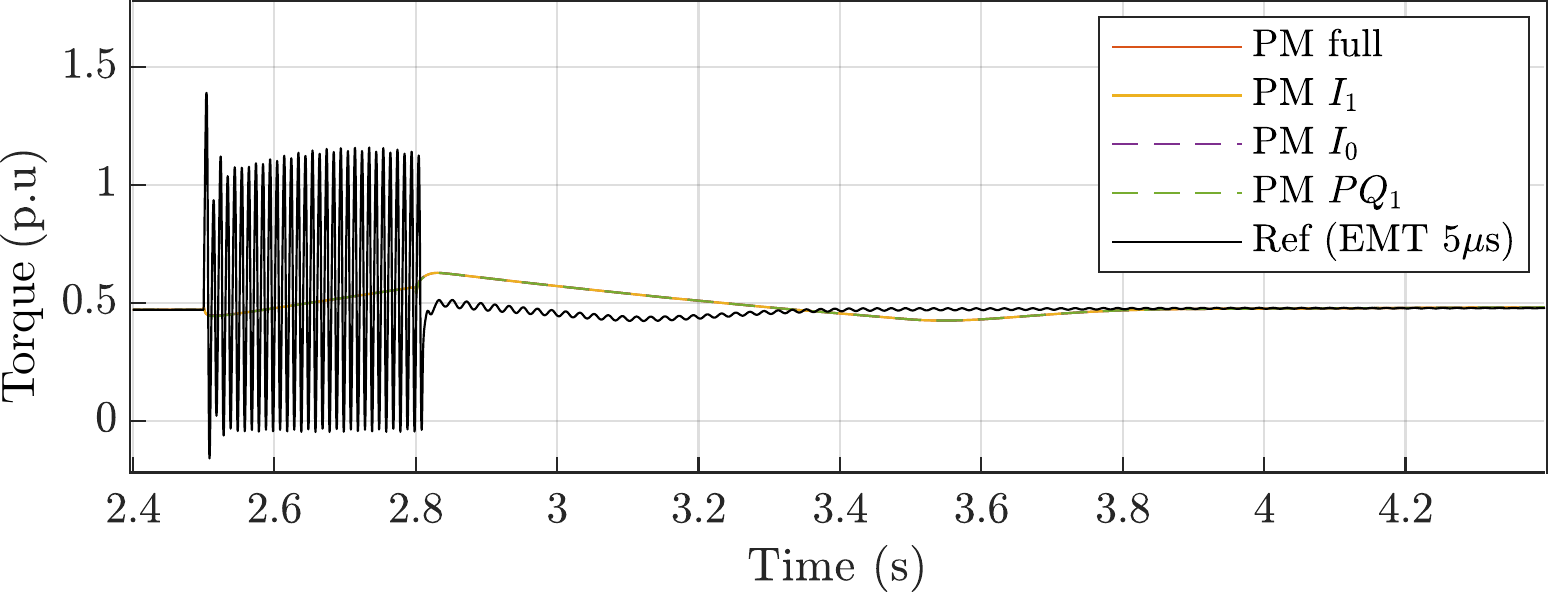}
	\vspace{-3pt}
	\caption{Test 4 -- $T_e$ simulated by each model at $350 \,\mu$s.}
	\label{fig_test4_Te_350us}
\end{figure}

During an asymmetric fault, EMT models show an oscillatory torque produced by the transient, which only the average value is captured by the PM models, as shown in Fig.~\ref{fig_test4_Te_350us}. Despite the higher error during the transient, the PM models match the EMT model closely after less than 1 second.

\begin{figure}[ht]
	\vspace{-3pt}
	\centering
	\includegraphics[width=0.46\textwidth]{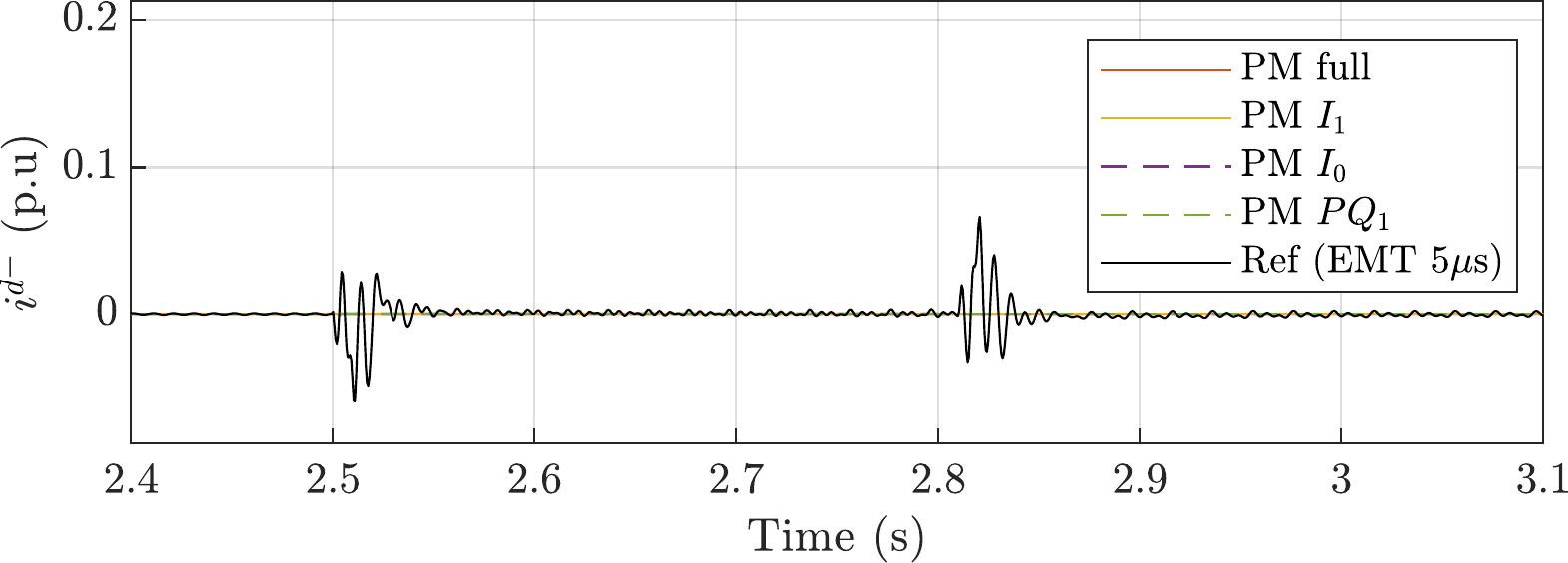}
	\vspace{-3pt}
	\caption{Test 4 -- $i^{d-}$ simulated by each model at $600 \,\mu$s.}
	\label{fig_test4_idn_350us}
\end{figure}

As the control strategy adopted for negative sequence control was $i_c^{q-\star} = 0$ and $i_c^{d-\star} = 0$, these variables were controlled to zero even during the transient in the PM models, as shown in Fig.~\ref{fig_test4_idn_350us}. The results with the PM models were more optimistic than the ones with EMT because the symmetric components in PM models are already available and do not need to be estimated as in EMT. This avoids the transient seen in Fig.~\ref{fig_test4_idn_350us} after the beginning and the end of the fault. 

\subsubsection{Harmonics}
In this test, a predefined harmonic content was injected into the system using harmonic current sources at $t \!=\! 0.5\,s$ and at $t \!=\! 0.6$~s the VSC active power setpoint was set to 50~MW. 
The harmonic orders were 2, 3, 4, 5, 7, 11 and 13, with magnitudes 1\%, 25\%, 2.5\%, 15\%, 7.5\%, 4\% and 2.5\%, respectively, with reference to a 30 MW, 10 MVar load.

\begin{figure}[ht]
	\vspace{-3pt}
	\centering
	\includegraphics[width=0.46\textwidth]{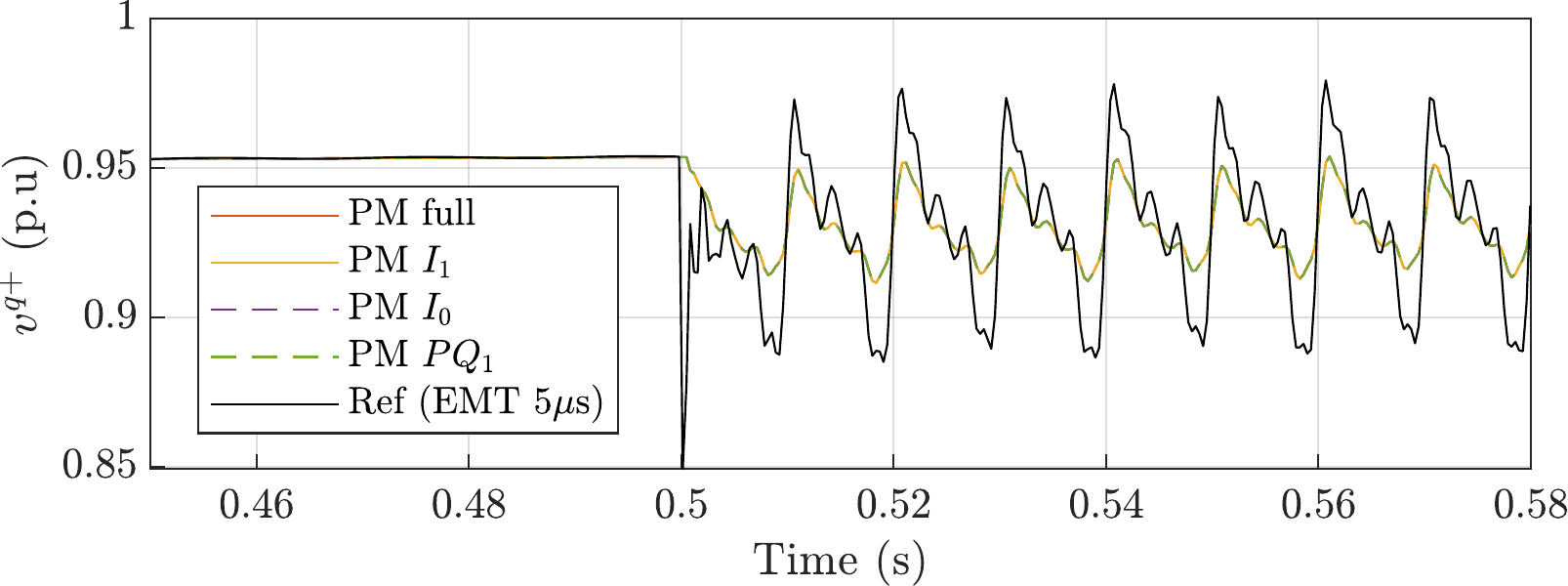}
	\vspace{-3pt}
	\caption{Test 5 -- $v^{q+}$ simulated by each model at $350 \,\mu$s.}
	\label{fig_test5_vqp_350us}
\end{figure}

Fig.~\ref{fig_test5_vqp_350us} presents $v_{q+}$ measured at the converter terminal after the injection of the harmonic components. As can be observed, the PM models, though not matching the EMT model, also revealed the harmonic corruption in the voltage. 

\begin{figure}[ht]
	\vspace{-3pt}
	\centering
	\includegraphics[width=0.46\textwidth]{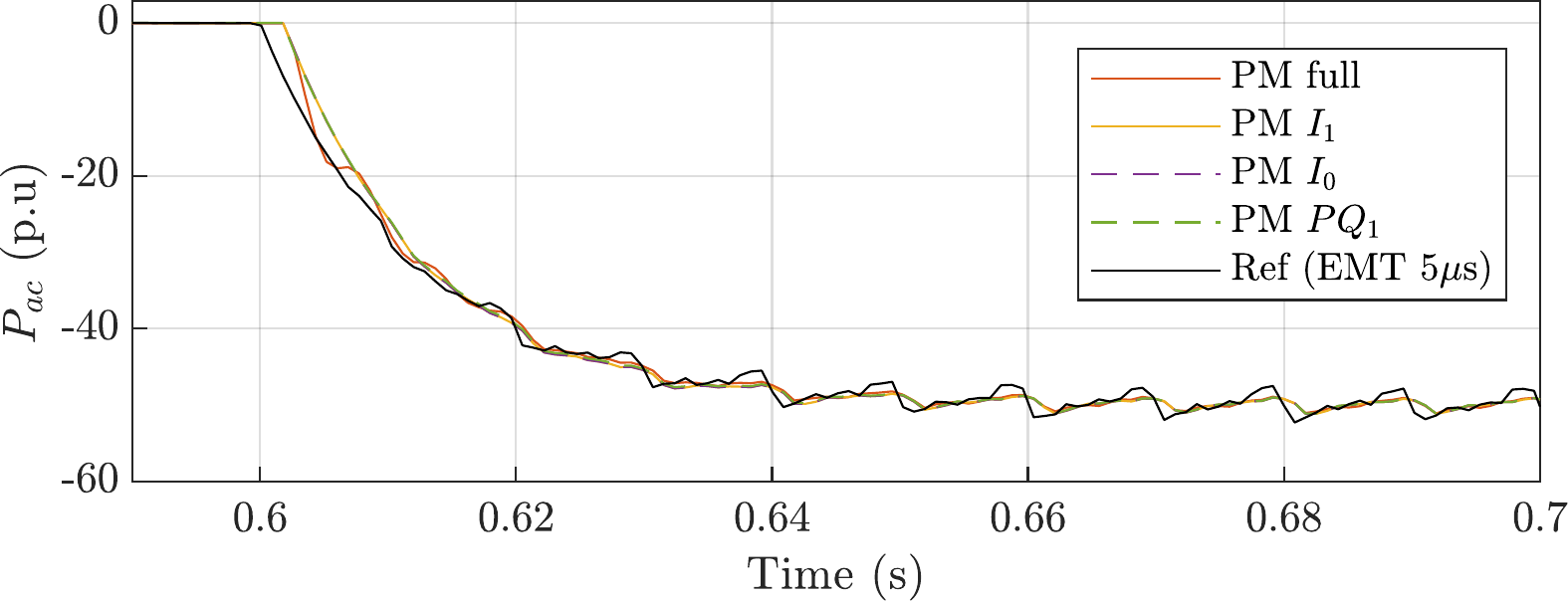}
	\vspace{-3pt}
	\caption{Test 5 -- $P_{ac}$ simulated by each model at $850 \,\mu$s.}
	\label{fig_test5_Pac_850us}
\end{figure}

\begin{figure}[ht]
	\vspace{-3pt}
	\centering
	\includegraphics[width=0.46\textwidth]{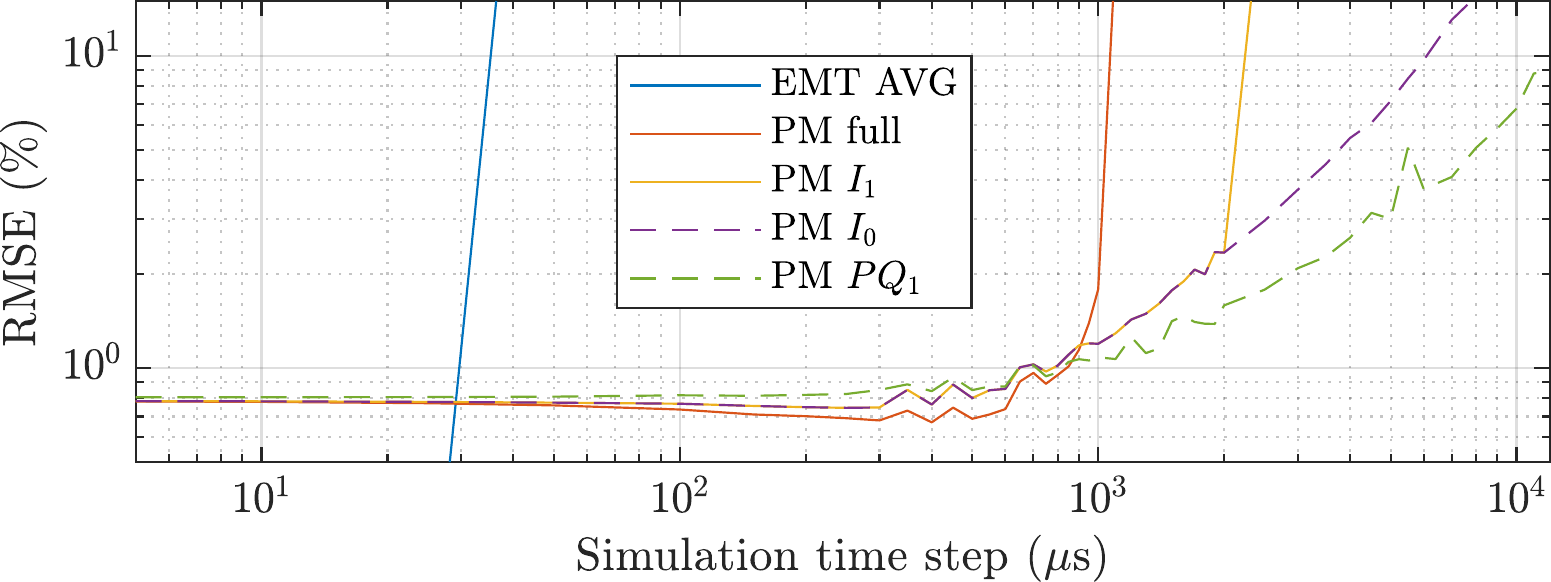}
	\vspace{-3pt}
	\caption{Test 5 -- $P_{ac}$ setpoint tracking RMSE simulated by each model at several time steps.}
	\label{fig_test5_Pac_RMSE}
\end{figure}

However, for variables with slower dynamics, the error of PM models was smaller, even with harmonics, as shown in Fig.~\ref{fig_test5_Pac_850us} and in Fig.~\ref{fig_test5_Pac_RMSE}.


\subsection{Large system}
A larger system composed of four synchronous generators and two VSCs was simulated. Frequency and voltage droops, LVRT and negative sequence control were enabled during all tests.


  \subsubsection{Asymmetrical fault}
\begin{figure}[ht]
	\vspace{-3pt}
	\centering
	\includegraphics[width=0.46\textwidth]{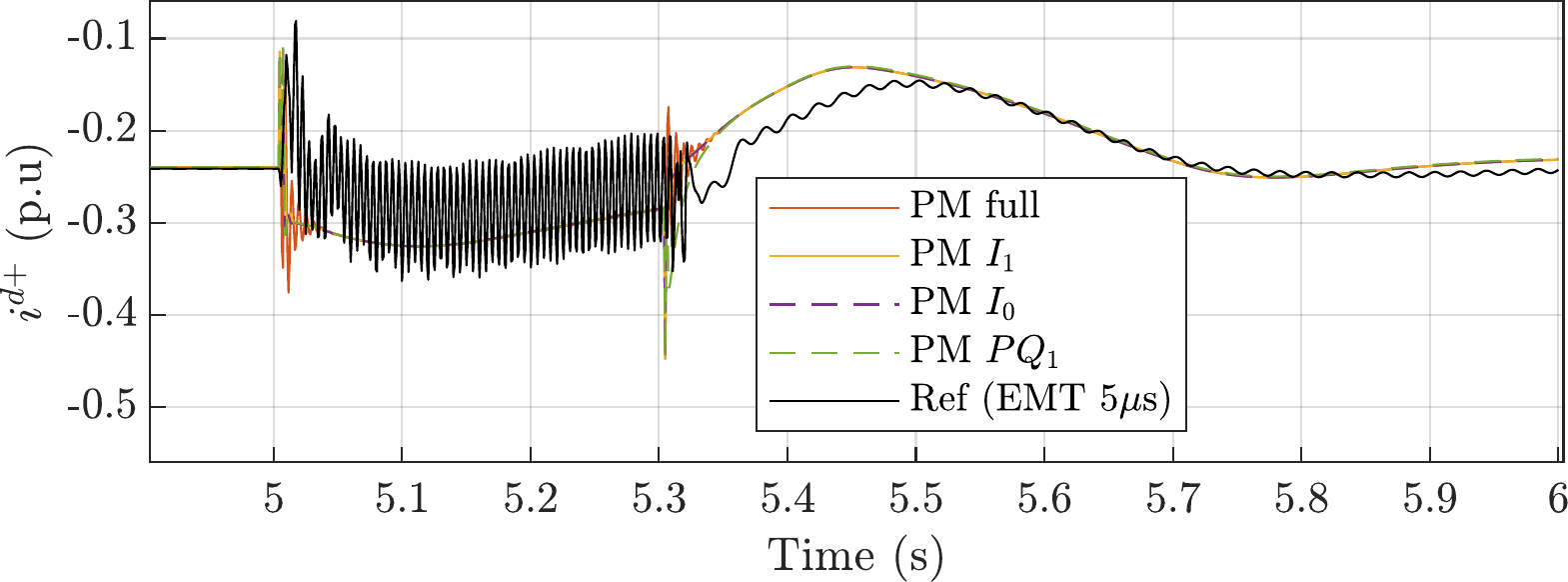}
	\vspace{-3pt}
	\caption{Test 1 large system -- $i^{d+}$ simulated by each model at $850 \,\mu$s.}
	\label{fig_large_test1_idp2_850us}
\end{figure}

In this test, an 1~$\Omega$ mono-phase fault at phase C was applied to the system at $t \!=\! 5.0\,s$, lasting for 300 ms.
Fig.~\ref{fig_large_test1_idp2_850us} shows VSC~2 $i^{d+}$ simulated at $850 \,\mu$s. 

\begin{figure}[ht]
	\vspace{-3pt}
	\centering
	\includegraphics[width=0.46\textwidth]{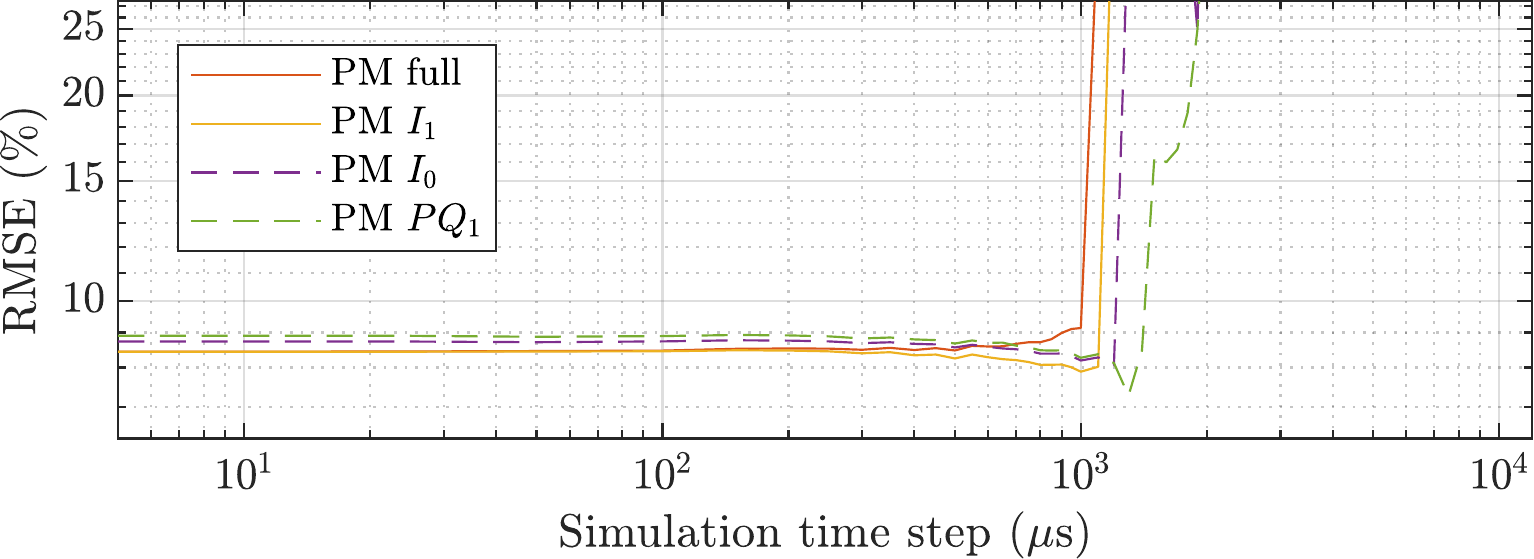}
	\vspace{-3pt}
	\caption{Test 1 large system -- $i^{d+}$ RMSE simulated by each model at several time steps.}
	\label{fig_large_test1_RMSE_idp}
\end{figure}

As can be observed in Fig.~\ref{fig_large_test1_idp2_850us}, during the transient, the PM models assume an average value in relation to the EMT simulation, and the difference between both model types reduces 300~ms after the end of the fault. In the large system, more elements and controllers interact through long transmission lines, which increase the transient response and wider the difference between EMT and PM models. However, even for the large system, the PM models were still able to track the average value of the EMT simulation. However, the fastest dynamics in the system are still the bottleneck to increasing the simulation time step. As shown in Fig.~\ref{fig_large_test1_RMSE_idp}, after the time step increases beyond 1~ms the PM models error increased dramatically.   

  \subsubsection{Loss of generation}
In this test, the generator G1 is disconnected from the system at $t \!=\! 5.0\,s$, producing a slow but large transient in the system.

\begin{figure}[ht]
	\vspace{-3pt}
	\centering
	\includegraphics[width=0.46\textwidth]{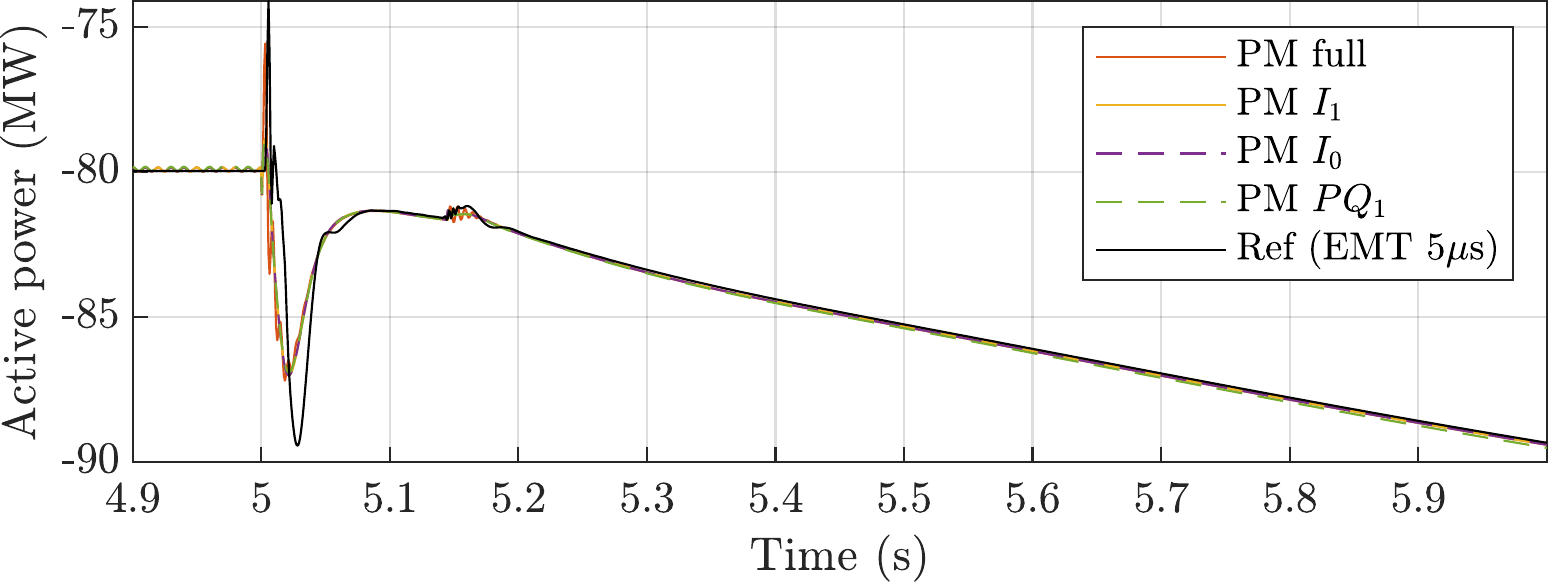}
	\vspace{-3pt}
	\caption{Test 2 large system -- VSC 1 $P_{ac}$ simulated by each model at $750 \,\mu$s.}
	\label{fig_large_test2_Pac1_750us}
\end{figure}

\begin{figure}[ht]
	\vspace{-3pt}
	\centering
	\includegraphics[width=0.46\textwidth]{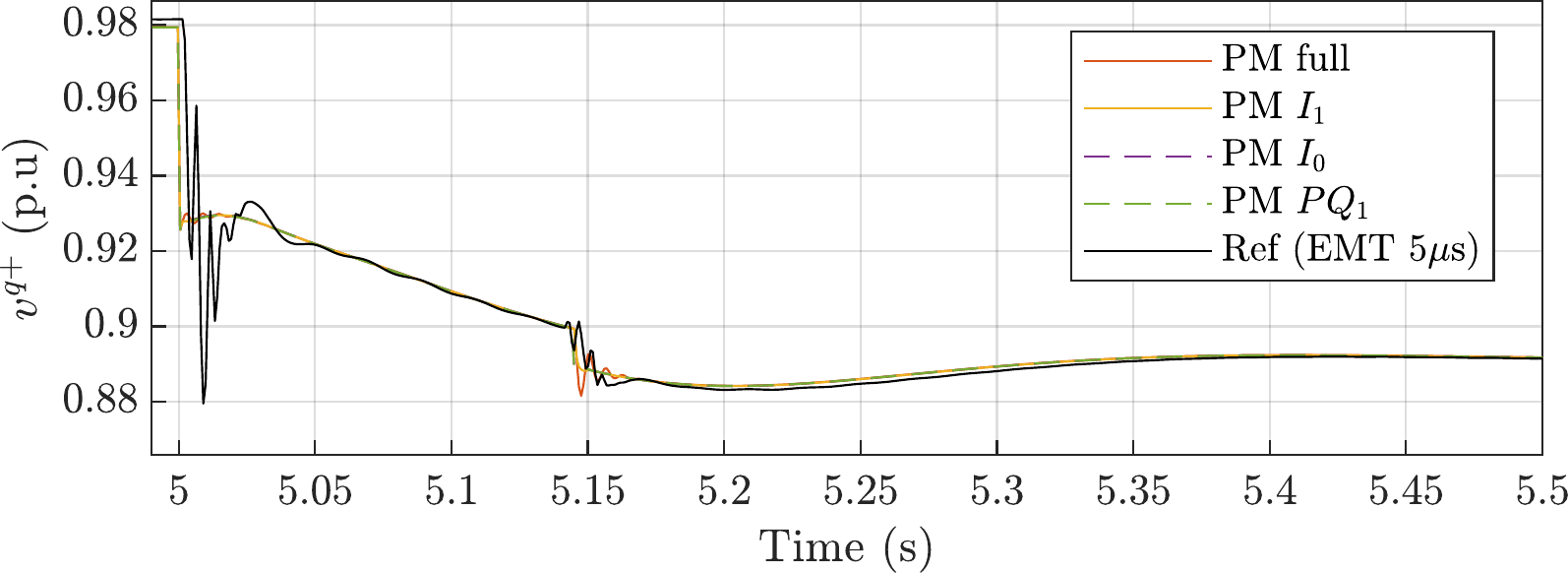}
	\vspace{-3pt}
	\caption{Test 2 large system -- VSC 1 $v^{q+}$ simulated by each model at $850 \,\mu$s.}
	\label{fig_large_test2_vqp2_850us}
\end{figure}

As the transient was electromechanical and did not affect the voltage significantly, the converter's PLLs remained synchronized, and the difference between PM and EMT models was observable only for about 100~ms, as shown in Fig.~\ref{fig_large_test2_Pac1_750us} and Fig.~\ref{fig_large_test2_vqp2_850us}.

  \subsubsection{Line outage}
In this test, a permanent 1~$\Omega$ mono-phase fault at phase C was applied to the line connecting bus 1 to bus 3. The line was isolated by two circuit breakers, the first opening at 200~ms and the second at 250~ms after the fault, producing a transient in the system.

\begin{figure}[ht]
	\vspace{-3pt}
	\centering
	\includegraphics[width=0.46\textwidth]{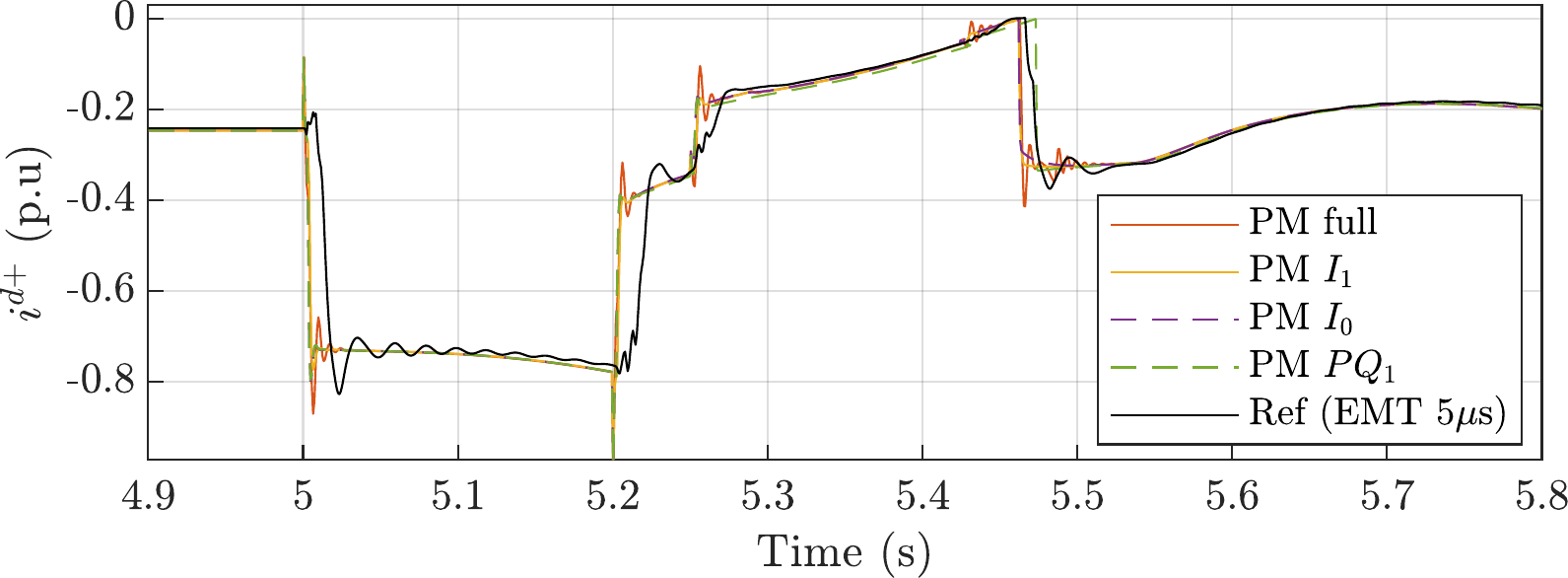}
	\vspace{-3pt}
	\caption{Test 3 large system -- VSC 1 $i^{d+}$ simulated by each model at $850 \,\mu$s.}
	\label{fig_large_test3_idp1_850us}
\end{figure}

\begin{figure}[ht]
	\vspace{-3pt}
	\centering
	\includegraphics[width=0.46\textwidth]{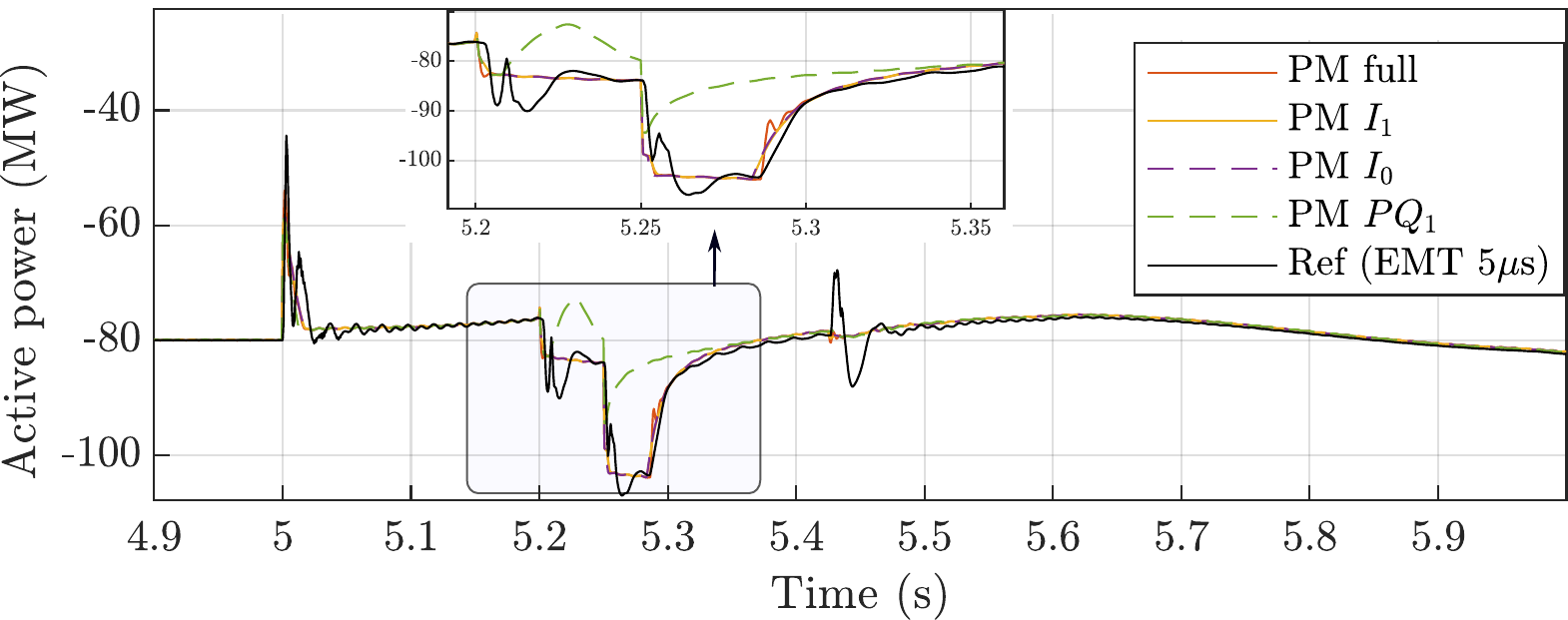}
	\vspace{-3pt}
	\caption{Test 3 large system -- VSC 1 $P_{ac}$ simulated by each model at $700 \,\mu$s.}
	\label{fig_large_test3_Pac1_700us_comzoom}
\end{figure}

After each transient (fault, first breaker opening and second breaker opening) it is possible to observe in Fig.~\ref{fig_large_test3_idp1_850us} and Fig.~\ref{fig_large_test3_Pac1_700us_comzoom} that the difference between EMT and most PM models was high only for about 30~ms. The same is not true for the most approximated PM $PQ_1$, which took about 100~ms to match the EMT model after each event.

\begin{figure}[ht]
	\vspace{-3pt}
	\centering
	\includegraphics[width=0.46\textwidth]{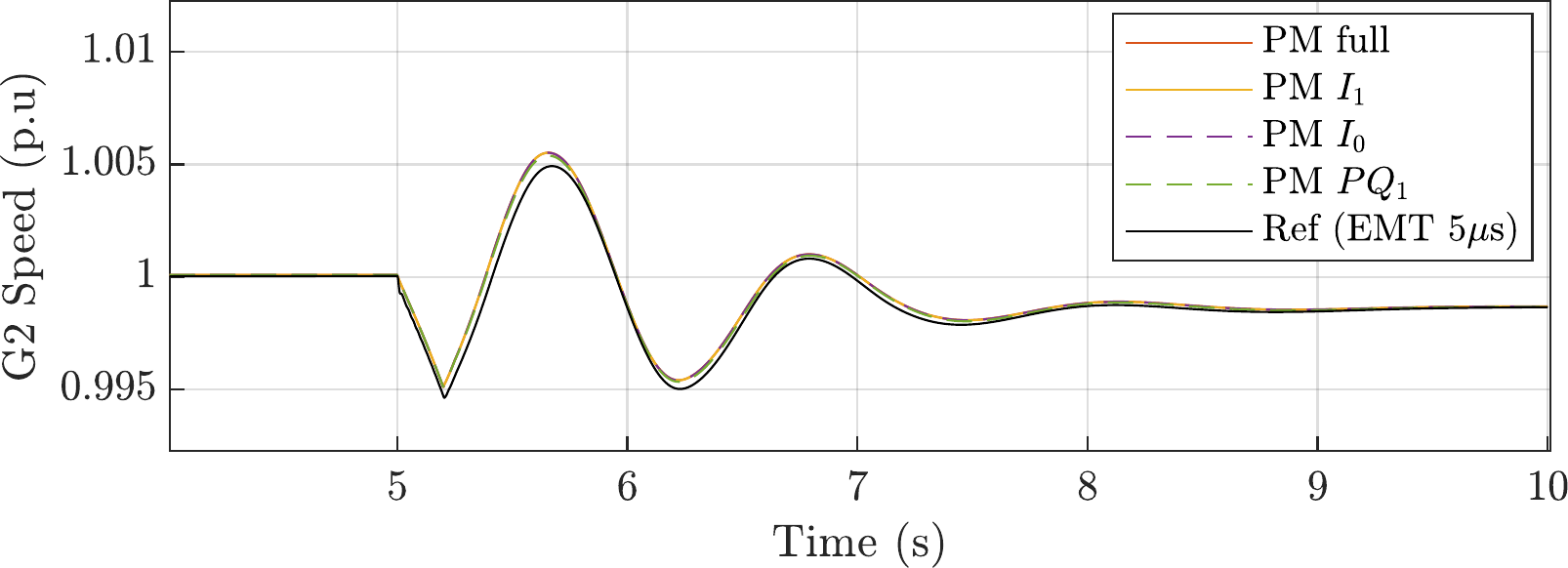}
	\vspace{-3pt}
	\caption{Test 3 large system -- SG 3 speed simulated by each model at $750 \,\mu$s.}
	\label{fig_large_test3_wm3_750us}
\end{figure}

\begin{figure}[ht]
	\vspace{-3pt}
	\centering
	\includegraphics[width=0.46\textwidth]{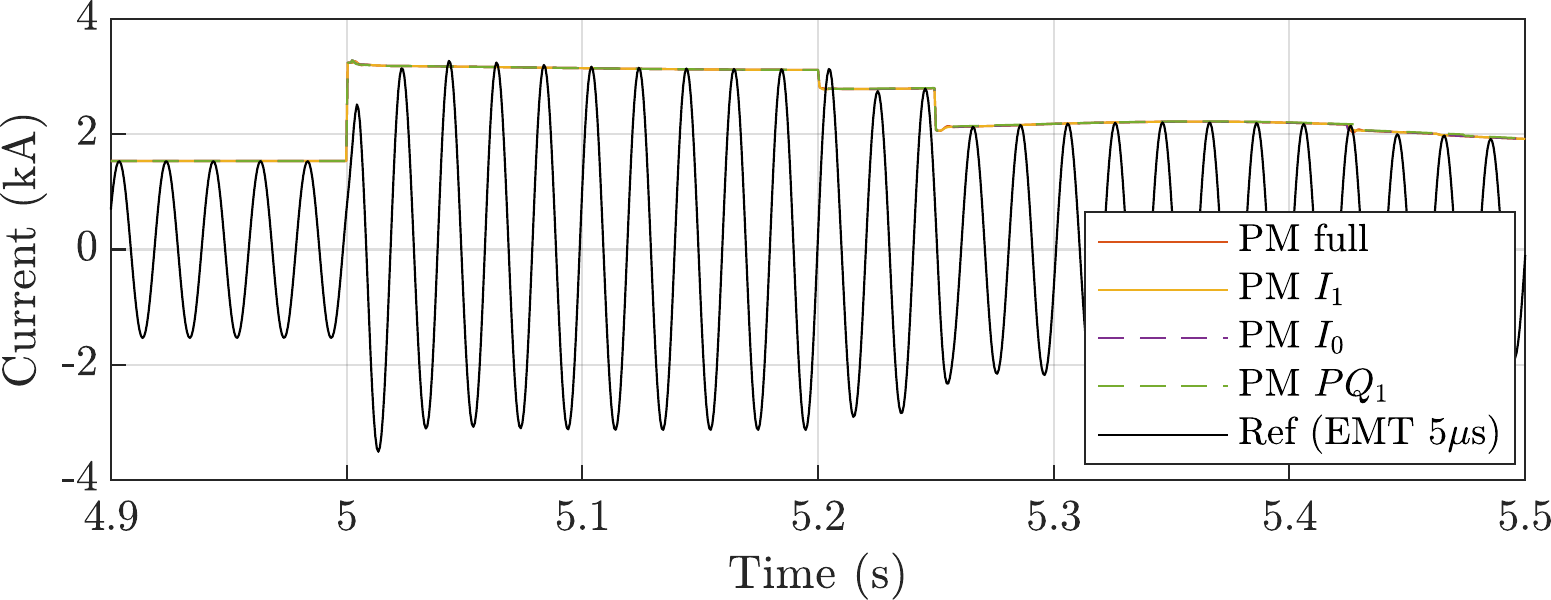}
	\vspace{-3pt}
	\caption{Test 3 large system -- SG 2 $I_c$ simulated by each model at $650 \,\mu$s.}
	\label{fig_test4_Ic_650us}
\end{figure}

From the SGs side, a small deviation could be observed in the speed (Fig.~\ref{fig_large_test3_wm3_750us}) while the current was precisely tracked after one cycle (Fig.~\ref{fig_test4_Ic_650us}).

\section{Discussion and Guidelines}\label{sec_guidelines}
\begin{figure}[ht]
	\vspace{-3pt}
	\centering
	\includegraphics[width=0.45\textwidth]{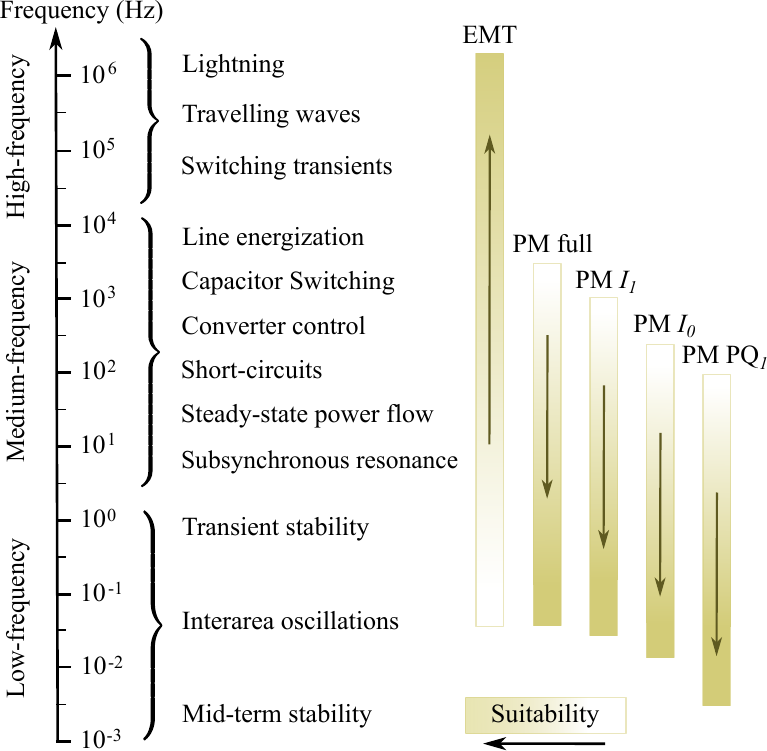}
	\vspace{-6pt}
	\caption{Suitability of each model. Frequency ranges are approximated for illustrative purposes \cite{Henschel_1999,Watson_2018}. \vspace{-12pt}}
	\label{fig_guidelines_summary}
\end{figure}

Based on the tests performed and our experience with EMT and PM models, the following discussion and guidelines follow. They are also summarized in Fig.~\ref{fig_guidelines_summary}. 

The major advantage of PM models is increasing the simulation time step and reducing the simulation time. If a proper time step is chosen, the simulation speed can be increased in one or two orders of magnitude, transforming hours of simulation to minutes. This allows larger models to be simulated or a higher number of tests to be performed in a feasible time. Another advantage of PM models is a more straightforward interpretation of the results. Analysing current and voltage phasors is more simple than voltage and current waveforms. Thus, using PM models can also contribute to increase the level of understanding of the simulation and consequently speed up the post-simulation time, where the analysis and interpretation of the results are performed.    

However, one must carefully analyse if the chosen PM model is able to represent well the set of studies to be performed. The results shown in the previous sections show that tests involving electromechanical transients and non-solid symmetrical faults are well represented by PM models. However, asymmetrical faults and symmetrical faults that drops voltage near to zero dramatically influence converters' PLLs. As all PM models consider an ideal PLL, their results might be mistakenly optimistic in these cases. In all cases, the PM models take about 100~ms to closely track EMT models (this time vary depending on the system equivalent X/R ratio). Thus, in the simulated systems, transients faster than 100~ms might be overlooked by PM models. 

Another important aspect to be considered in both EMT and PM simulation is the choice of the simulation time step. Choosing an arbitrary low time step may lead to unnecessary slow simulation, with much data being produced. Choosing an arbitrary high time step might lead to inaccurate results or numerical instability. Based on the results of the previous section, we recommend that the simulation time step should be at least between five and ten times smaller than the smallest time constant being simulated. In theory, it could be up to the Nyquist frequency, which is half of the shortest time constant, but this would be too close from an erroneous simulation. This guideline agrees with past recommendations found in the literature \cite{Dommel_1969,de_Siqueira_2014}. Based on this rule, PM models can be even used to perform studies with harmonic content, provided that the greatest harmonic frequency is well represented by the simulation time step.

Among the PM models discussed in this paper, PM $I_1$ and $I_0$ offer the best trade-off between simulation time step and accuracy. Their drawback is that the converter's current control is approximated (PM $I_1$) or neglected (PM $I_0$), assuming that the current reference is always correctly tracked. However, if the converter's internal variables are not the focus of the study, these models can significantly increase the simulation speed providing good precision at the same time.

\section{Conclusion}\label{sec_conclusion}
This paper reviewed EMT and PM models currently used to simulate VSC in EMT and Transient Stability studies. Several modelling guidelines and suitability analyses were provided based on a comprehensive comparative study amongst the models. We addressed a gap related to the suitability of PM models in studies where the boundary between electromagnetic and electromechanical domains were overlapped. An insightful analysis of the adequate simulation time step for each model and study was provided. The recommendations were based on a systematic comparison amongst the models in several different scenarios. 

From the tests performed, it could be observed that the major advantage of PM models is the possibility to increase the simulation time step and reduce the simulation time. Another advantage is a more straightforward interpretation of the results when compared to detailed waveforms in EMT.
However, the set of studies to be performed by PM models must be carefully analysed. As the PM models consider some idealistic conditions, e.g an ideal PLL, their results might be mistakenly optimistic in cases involving large transients.

The recommendations provided aim to contribute to the discussion about the most suitable models of modern power grids, where the widespread of power electronic interfaced technologies and the reduced inertia are making the grid's dynamic behaviour to become progressively faster.

\appendices
\section{Test systems parameters}

The SGs parameters of the large system are summarized in Table~\ref{tab_SG_large_system}. The SG parameters of the small system are equal to G1 in the large system, but with nominal power equal to 400~MVA. The transmission lines parameters at 50~Hz are: $R_1 = 0.121 \,\Omega$/km, $R_0 = 0.446 \,\Omega$/km, $L_1 = \,1.33$mH/km, $L_0 \!=\! \, 3.73$mH/km, $C_1 \!=\! \,8.762$ nF/km and $C_0 \!=\! \,6.373$ nF/km.

\vspace{-3pt}

\begin{table}[h]
	\centering
		\caption{Large system parameters - Synchronous generators.}
	\begin{tabular}{@{\hspace{1\tabcolsep}}lllll@{\hspace{1\tabcolsep}}}
		\toprule
		& $G0$    & $G1$    & $G2$    & $G3$    \\ \midrule
		&       &       &       &       \\
		$S_n$ (MVA)  & 1000  & 700   & 500   & 500   \\
		$V_n$ (kV)   & 22.0  & 22.0  & 22.0  & 22.0  \\
		$X_d$ (pu)   & 2.00  & 1.25  & 1.667 & 1.25  \\
		$X_q$ (pu)   & 1.80  & 1.00  & 1.125 & 1.00  \\
		$X_d'$ (pu)  & 0.35  & 0.333 & 0.25  & 0.333 \\
		$X_d''$ (pu) & 0.25  & 0.292 & 0.233 & 0.292 \\
		$X_q''$ (pu) & 0.30  & 0.292 & 0.225 & 0.292 \\
		$X_l$ (pu)   & 0.15  & 0.15  & 0.15  & 0.15  \\
		$R_s$ (pu)   & 0.01  & 0.01  & 0.01  & 0.01  \\
		$T_{do}'$ (s)  & 4.485 & 5.00  & 6.00  & 5.00  \\
		$T_{do}''$ (s) & 0.068 & 0.002 & 0.002 & 0.002 \\
		$T_{qo}''$ (s) & 0.10  & 0.002 & 0.002 & 0.002 \\
		$H$ (s)     & 6.0   & 5.0   & 3.0   & 5.0 \\ \bottomrule
		\vspace{-18pt}
	\end{tabular}
\label{tab_SG_large_system}
\end{table}





\ifCLASSOPTIONcaptionsoff
  \newpage
\fi


\bibliographystyle{IEEEtran}
\bibliography{References}

%








\end{document}